\documentclass[
reprint,
superscriptaddress,
groupedaddress,
amsmath, 
amssymb,
aps,
prd,
floatfix
]{revtex4-2}

\usepackage{graphicx} 
\usepackage[caption=false]{subfig}
\usepackage{dcolumn} 
\usepackage{bm} 
\usepackage{xspace}
\usepackage{pifont}
\usepackage{braket}
\usepackage{enumerate}
\usepackage{booktabs}

\usepackage[dvipsnames]{xcolor}
\definecolor{azulclaro}{HTML}{14AAD6}
\definecolor{azuloscuro}{HTML}{0057CD}
\usepackage[colorlinks=true, linkcolor=blue, citecolor=blue, urlcolor=blue]{hyperref}	
\hypersetup{
	pdftitle={Uncovering subdominant multipole asymmetries in binary black-hole mergers},
	pdfauthor={Jannik Mielke, Angela Borchers, Frank Ohme}
}

\graphicspath{{./figures/}}

\newcommand{\AEI}{Max Planck Institute for Gravitational Physics (Albert Einstein Institute), D-30167 Hannover, Germany}
\newcommand{\LUH}{Leibniz Universität Hannover, D-30167 Hannover, Germany}

\begin{document}

\title{Uncovering subdominant multipole asymmetries in binary black-hole mergers}

\author{Jannik Mielke}
\email{jannik.mielke@aei.mpg.de}
\author{Angela Borchers}
\author{Frank Ohme}

\affiliation{\AEI}
\affiliation{\LUH}

\date{\today}

\begin{abstract}
In dynamically formed binaries, the spins of the black holes tend to be misaligned with the system’s orbital angular momentum. This causes the spins to precess and leads to an asymmetric emission of gravitational waves. The resulting gravitational-wave multipole asymmetries directly source the recoil of the remnant black hole and are the critical element in fully describing precession. Recoil and precession are of significant astrophysical importance, but multipole asymmetries contribute only minimally to the overall signal strength. Consequently, most current gravitational-wave models either do not incorporate asymmetries at all, or only consider the dominant ones. Here we highlight the importance of subdominant multipole asymmetries for an accurate recoil velocity calculation and discuss their detectability with third generation detectors. Neglecting subdominant asymmetries leads to velocity differences of up to 210 km/s and can, in particular, introduce systematic biases in the inference of masses and the spin geometry. We further discuss universal characteristics of subdominant multipole asymmetries in order to prepare the ground for potential future asymmetry models. In the inspiral regime, the average antisymmetric frequencies can be described by a multiple of the orbital frequency. During ringdown, however, they become equal to their corresponding symmetric frequencies.
\end{abstract}

\maketitle

\section{Introduction}
\label{sec:introduction} 

General relativity (GR) predicts that gravitational waves (GWs) have a quadrupolar nature. Accordingly, when decomposing the waveform in a $s=-2$ spin-weighted spherical harmonic (SWSH) basis, the $(\ell,m)=(2,\pm 2)$ multipoles dominate the radiation from compact binary coalescences (CBCs). Hence, most of the scientific output from the 218 detected CBC signals by the LIGO-Virgo-Kagra Collaboration \cite{LVK_2019_GWTC1, LVK_2021_GWTC2, LVK_2023_GWTC3, LVK_2025_GWTC4} arises from accurately modeling these dominant multipoles.

However, GWs exhibit additional structure beyond the leading quadrupole, including higher-order multipoles and asymmetries between the $+m$ and $-m$ multipoles. In the cases of GW190412 \cite{LVK_2020_GW190412, Bustillo_2022_GW190412}, GW190814 \cite{LVK_2020_GW190814} and GW241011 \cite{LVK_2025_GW241011} evidence for higher-order multipoles has been reported. For GW200129\_065458, the presence of a dominant multipole asymmetry played a key role in the observation of the remnant's recoil, orbital and spin precession \cite{Hannam_2022_GW200129, Varma_2022_GW200129, Kolitsidou_2024_AsymSpinMeasure, Borchers_2024_ObsPrecKick}. 

In precessing binaries, the precession itself and the largest contribution to the recoil, also known as the ``kick'', arise from the components of the binary's spins lying in the orbital plane \cite{Apostolatos_1994_Precession, Kidder_1995_Mark52PN, Pretorius_2007_BBHCoalescence}. These spin components lead to an asymmetric emission of gravitational waves \cite{Bruegmann_2007_Superkicks, Campanelli_2007_superkick, Gonzalez_2007_superkick}. In particular, asymmetries in the multipolar structure of the radiation generate a net flux of linear momentum out of the orbital plane, causing the remnant black hole (BH) to recoil in the opposite direction of this flux \cite{Ruiz_2007_MomentaFluxes, Mielke_2025_kickasym}.

Observations of precession and kicks with GW detectors are astrophysically highly significant. Spin misalignment can help to reveal formation channels \cite{Gerosa_2013_formationchannel1, Vitale_2017_formationchannel2,Farr_2018_formationchannel3, Mapelli_2021_formationchannel4}, while large kick velocities may eject remnants from their host environments. These kicks limit the possibility of remnants undergoing further mergers, impacting the binary black hole (BBH) merger rate \cite{Gerosa_2021_popreview, Mahapatra_2021_HierachicalMergers} and the distributions of BH masses and spins \cite{AraujoAlvarez_2024_HierachicalGW190521, Mahapatra_2024_Genealogy, Borchers_2025_HierMerg}. In addition, accurate kick estimates, particularly for out-of-plane recoils, are crucial for interpreting potential multi-messenger observations of BBH mergers \cite{Leong_2025_KickMultiMessenger}, including scenarios such as those proposed in \cite{Graham_2022_BBHcounterparts}.

The GW emission from precessing binaries on quasi-circular orbits is fully described by seven intrinsic parameters. These are the mass ratio $q = m_1/m_2 \geq 1$ and the dimensionless spin vectors $\vec{\chi}_1$ and $\vec{\chi}_2$. 

To address the complete two-body problem in GR, one would in principle need to account for orbital eccentricity. In the following, however, we neglect this effect, since eccentricity decreases towards the merger, whereas the asymmetry in the $\pm m$ multipoles relevant to our study increases. Nevertheless, a recent study emphasizes the importance of accurately understanding multipole asymmetries in both quasi-circular and eccentric orbital configurations, since a consistent definition of eccentricity in eccentric and precessing systems is only possible with a correct modeling of the dominant asymmetry \cite{Shaikh_2025_EccDef}. In particular, the frequency difference between the $(2,2)$ and $(2,-2)$ multipoles in the coprecessing frame is the key for their waveform-based definition.

Currently, only a few waveform models incorporate multipole asymmetries. \texttt{IMRPhenomXO4a} \cite{Thompson_2024_IMRPhenomXO4a, Ghosh_2024_ModeAsymmetryXO4a} and \texttt{IMRPhenomXPNR} \cite{Hamilton_2025_XPNR} implement the dominant asymmetry. More recently, asymmetric contributions to the $\ell=m \leq 4$ multipoles have been incorporated in \texttt{SEOBNRv5PHM}$_{\text{w/asym}}$ \cite{Estelles_2025_EOBasym}. While these models accurately capture the physics of the asymmetries they include, they do not account for all subdominant contributions. In contrast, the \texttt{NRSur7dq4} \cite{Varma_2019_NRSur7dq4} model includes all asymmetries up to $\ell=4$. However, as a data-driven model, it is not explicitly informed by the underlying physical equations, but instead learns an optimal representation directly from the numerical-relativity (NR) data.
 
The construction of effective one-body (EOB) and phenomenological models relies on a combination of analytical frameworks, most prominently post-Newtonian (PN) expansions and BH perturbation theory formulated in terms of quasi-normal modes (QNM), with phenomenological modeling strategies and state-of-the-art NR simulations. Models are often constructed using simplifying frames based on waveform quantities related to orbital parameters. For example, a coprecessing frame \cite{Schmidt_2011_QAframe, Boyle_2011_CoprFrame, Boyle_2014_PrecessingBHOperators} can be used, in which the coordinates are chosen such that the orbital angular momentum remains aligned with the $z$-axis at all times. A corotating frame extends this definition by tracking the separation vector between the two BHs along the $x$-axis. For perturbation theory, the most appropriate frame to work with is the super rest frame \cite{Leaver_1985_QNManalysis, Moreschi_1988_supercentre, Moreschi_1998_restframe, Zertuche_2022_BMSFrameQNM}. This is the center-of-mass frame of the remnant, with the remnant's spin aligned to the $z$-axis and supertranslation freedoms fixed. 

In Sec.~\ref{sec:importance}, we explain the motivation behind investigating the very small effects of subdominant asymmetries. We discuss phenomenological observations in NR data and relate them to concepts in PN and perturbation theory in Sec.~\ref{sec:phenom}.

\section{Importance of subdominant multipole asymmetries}
\label{sec:importance}

In this section, we evaluate the importance of subdominant multipole asymmetries in fully characterizing out-of-plane kicks and their effect on parameter estimation. To demonstrate the latter, we perform Bayesian inference on synthetic signals injected into a future ground-based detector.

\subsection{Preliminaries}
\label{sec:preliminaries}

\begin{figure}[tb]
	\includegraphics[width=\linewidth]{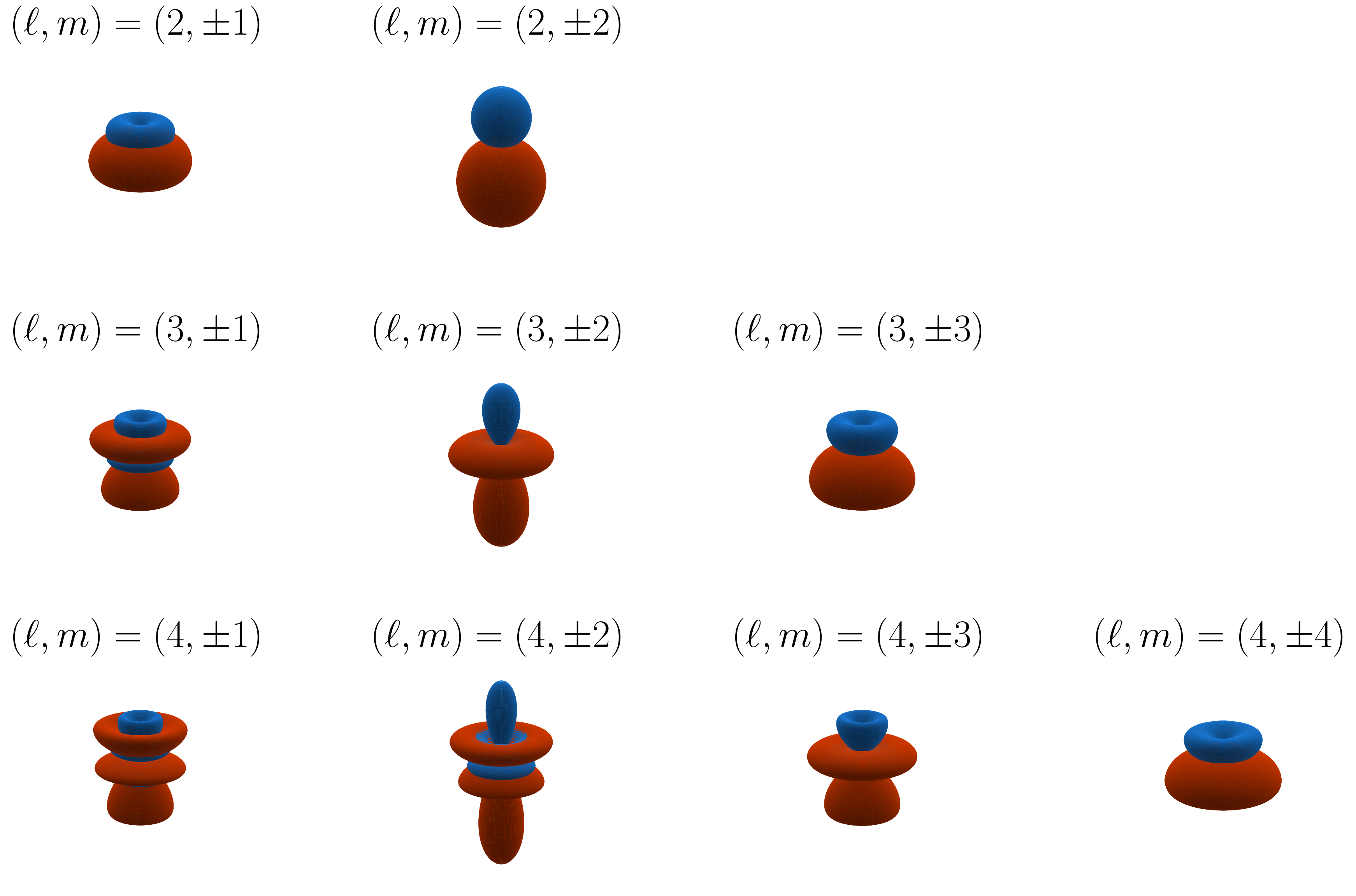}
	\caption{SWSH surface illustration. The radial distance from the origin encodes the magnitude $\left| {}^{-2}Y_{\ell m}\right|$ scaled with arbitrary amplitudes such that $\left|h_{\ell,+m}\right| \big/ \left|h_{\ell,-m}\right| = 2/3$ in order to illustrate an enlarged multipole asymmetry. Blue surfaces correspond to $m>0$, whereas red surfaces denote $m<0$.}
	\label{fig:swsh}
\end{figure}

The waveform of precessing systems is typically expressed as a decomposition in a $s=-2$ SWSH basis \cite{Goldberg_1967_delSWSH, Thorne_1980_Multipole},
\begin{equation}
	\label{eq:strain_decomposition}
	h(t, r, \theta, \varphi) =\sum_{\ell = 2}^{\infty} \sum_{ m = - \ell}^{\ell} {}^{-2}Y_{\ell, m}(\theta, \varphi)\,h_{\ell, m}(t, r) \, ,
\end{equation}
where the multipoles $h_{\ell,m}$ depend on all intrinsic source parameters. Fig.~\ref{fig:swsh} illustrates the ${}^{-2}Y_{\ell, m}$ basis weighted with arbitrary values for positive and negative $m$-multipoles.

For non-precessing binaries, the dynamics are confined to a fixed orbital plane, which induces a reflection symmetry in the gravitational radiation. As a consequence, the waveform multipoles satisfy
\begin{equation}
	\label{eq:mode_symmetry}
	h_{\ell,-m} =(-1)^{\ell}h_{\ell,m}^*\,.
\end{equation} 

Precessing binaries break this symmetry. To describe the multipole asymmetry one defines symmetric (``$+$'') and antisymmetric (``$-$'') multipoles as
\begin{equation}
	\label{eq:plus_minus_wf_def}
	h^{\pm}_{\ell,m} = \frac{1}{2} \left[h_{\ell, m} \pm (-1)^\ell h^*_{\ell,-m}\right]\,.
\end{equation} 

We further define the $+/-$ amplitude $|h^\pm_{\ell,m}| = a^\pm_{\ell,m}$, the phase $\arg(h^\pm_{\ell,m}) = \phi^\pm_{\ell,m}$, and frequency $\omega_{\ell,m}^\pm = \dot{\phi}^\pm_{\ell,m}$. Since the majority of the GW energy is emitted in the $\ell=2$, $m=\pm2$ multipoles, the quantity $h_{2,2}^-$ is referred to as the dominant antisymmetric waveform, whereas all other combinations of $\ell$ and $m$ are associated with subdominant multipole asymmetries.

\subsection{Complete characterization of the out-of-plane kick}
\label{sec:kick} 

\begin{figure}[tb]
	\includegraphics[width=\linewidth]{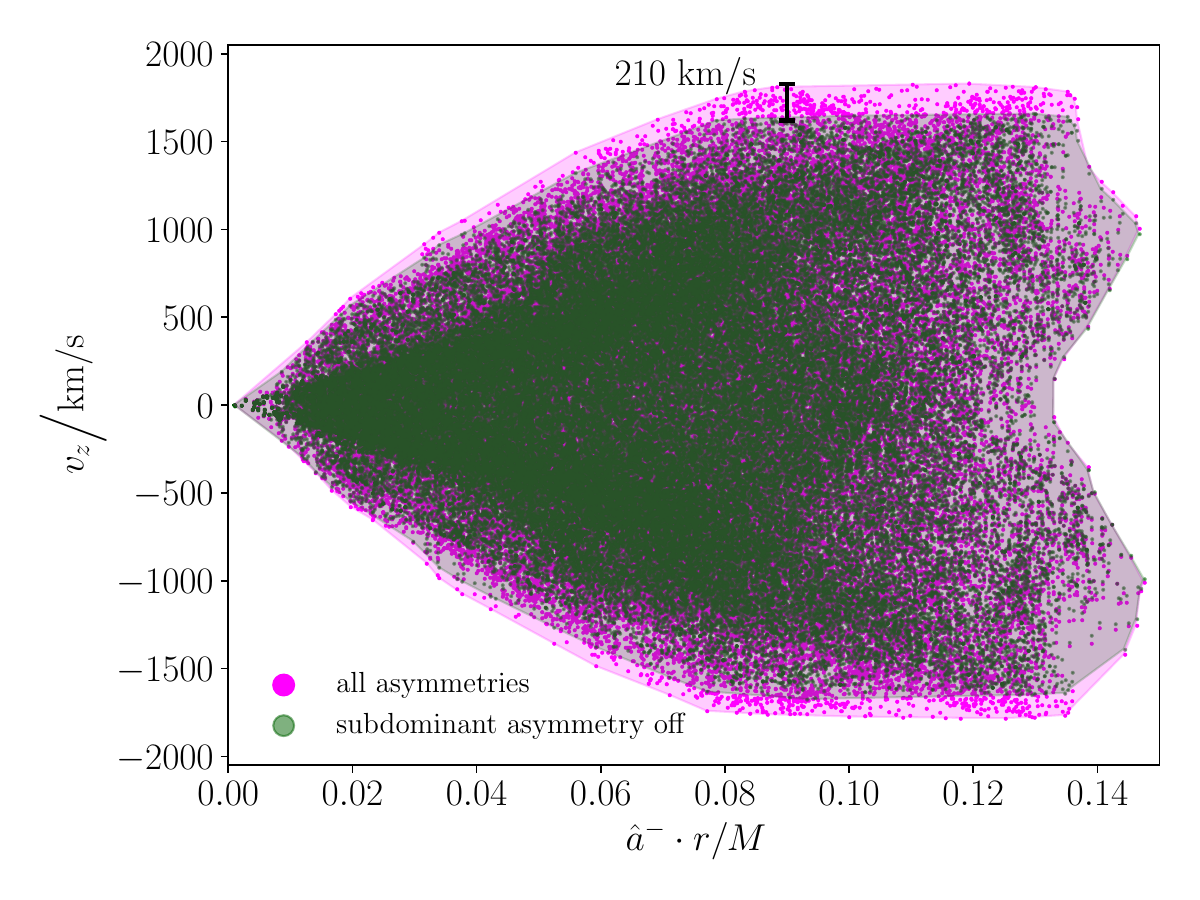}
	\caption{Spin–asymmetry–kick correlation, showing the difference between including only dominant multipole asymmetries and including all multipole asymmetries in the kick calculation. Configurations with $q=3$, spin magnitude of $\left|\vec\chi_i\right| = 0.8$ and varying spin directions were chosen. Waveforms were generated using \texttt{NRSur7dq4} \cite{Varma_2019_NRSur7dq4}. Kick computations and coprecessing-frame transformations were performed with \texttt{scri} \cite{Boyle_2020_scri}.}
	\label{fig:sak_q3_2spin}
\end{figure} 

\begin{figure*}[tb]
	\includegraphics[width=\linewidth]{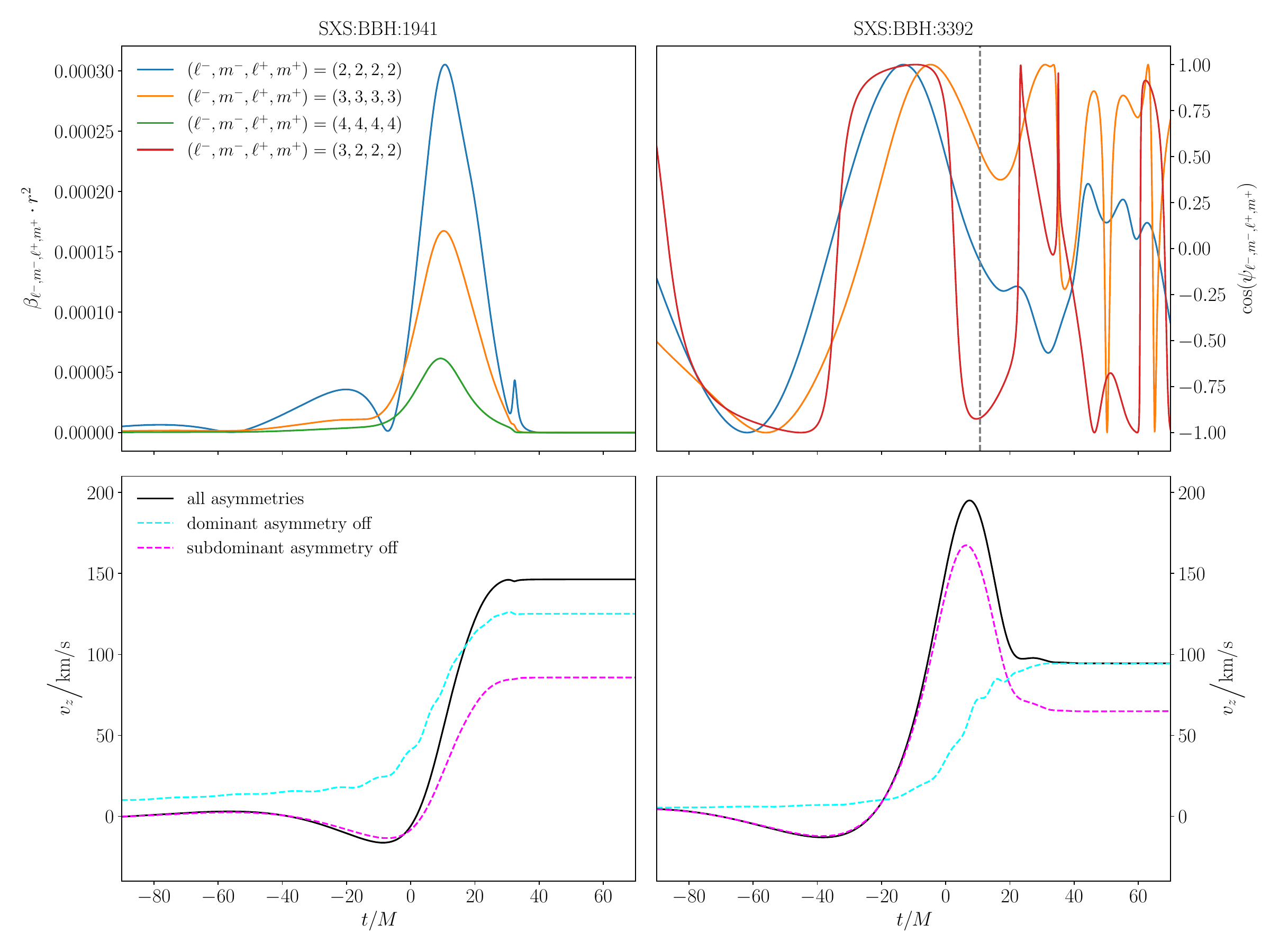}
	\caption{Influence of subdominant multipole asymmetries on the kick velocity. The left panel illustrates a case with large subdominant kick amplitudes $\beta$, while the right panel depicts a scenario in which the contribution of the dominant multipole asymmetry to the kick is almost negligible. This occurs because, at the time when most of the linear momentum is radiated (dashed gray line), the dominant antisymmetric and symmetric waveforms are nearly perpendicular, resulting in a cosine value of the phase difference $\psi$ close to zero. The lower panel provides the cumulative kick velocity calculated with all asymmetries, or when the dominant or subdominant asymmetry is deactivated.}
	\label{fig:subdom_kick}
\end{figure*} 

A fundamental characteristic of dominant and subdominant multipole asymmetries is that they describe the GW's linear momentum flux above and below the orbital plane of the binary. Due to momentum conservation, the remnant BH receives a kick in the opposite direction to the linear momentum flux given by \cite{Misner_1973_MTWGravitation, Thorne_1980_Multipole}
\begin{equation}
	\label{eq:dPidt_GR}
	\frac{\mathrm d P_z}{\mathrm d t} = \lim_{r \rightarrow \infty} \frac{r^2}{16\pi}\, \oint \mathrm d\Omega\, \cos\theta\,|\dot{h}|^2\,.
\end{equation}

We restrict our analysis to the kick along the $z$-direction. While an in-plane kick also exists \cite{Racine_2009_2PNKick, Borchers_2023_kickbasedGWtuning}, it is not affected by multipole asymmetries relative to the orbital plane, when we choose coordinates in a coprecessing frame \cite{Schmidt_2011_QAframe, Boyle_2011_CoprFrame, Boyle_2014_PrecessingBHOperators}. 

In this study, we focus on the impact of subdominant multipole asymmetry on the final kick velocity. We start by using a phenomenological tool from Sec.~III of our previous study \cite{Mielke_2025_kickasym}, the spin-asymmetry-kick correlation plots. These are distributions of various BBH configurations in the plane spanned by their out-of-plane kick and the maximum value of the antisymmetric amplitude $\hat{a}^- = \text{max}_t\left(\left|\sum_{\ell,m} h^-_{\ell,m}\right|\right)$. The kick magnitudes we will consider may lie within the uncertainty ranges of remnant or waveform models, yet the trends we report are expected to remain robust with increasing model accuracy.

In Fig.~\ref{fig:sak_q3_2spin}, we show equal-spin magnitudes with $q=3$, as these configurations exhibit a pronounced effect when the subdominant antisymmetric waveform is set to zero. In addition to observations such as the dependence of the kick range on $\hat{a}^-$, which we already found in our previous study, a new observation is that the full range of kick values cannot be recovered when only the dominant multipole asymmetries are included in the kick calculation. For binaries with strong precession and large antisymmetric waveforms, differences of up to 210 km/s are possible. This maximal difference arises from the combined contribution of all subdominant multipole asymmetries, which exhibit the same spin–asymmetry–kick phenomenology but produce much smaller individual kick magnitudes. 

Performing the spin-asymmetry-kick correlation studies with different mass ratios shows that, in approximately 5 \% of configurations, the kick differs by more than 100 km/s due to missing subdominant multipole asymmetries. This magnitude is astrophysically relevant, especially for systems with kick velocities comparable to the escape velocities of their host environments, such as young star clusters, globular clusters, or other possible environments \cite{Merritt_2004_vesc, Gerosa_2021_vesc}. In these environments, even small variations in recoil velocity can determine whether the remnant is retained or ejected. Kicks influence the spin distributions of BHs in these systems. Therefore, precise kick estimates are essential for reliably predicting the resulting spin distributions in such environments \cite{Borchers_2025_HierMerg}.

As the mass ratio increases, the ratio between the maximal kick difference and the maximal possible kick range, considering all multipoles, also increases, indicating the relevance of subdominant multipole asymmetries for high-mass ratio configurations.

Boyle et al.~\cite{Boyle_2014_PrecessingBHOperators} and Ma et al.~\cite{Ma_2021_SuperkickGWs} already initiated an analysis of the kick velocity in terms of multipole asymmetries. In our previous work, we derived an algebraic expression for $\mathrm{d}P_z/\mathrm{d}t$ that depends explicitly on interactions between the symmetric and antisymmetric parts of the waveform (cf.~Eq.~(20) in \cite{Mielke_2025_kickasym} and note that there is the close relation to mirror asymmetries analyzed for instance in Refs.~\cite{Leong_2025_MirrorAsymmetry, Bustillo_2024_LVKMirrorAsymmetry}). The dominating part of the kick, 
\begin{equation}
\label{eq:kick_beta_psi}
	\frac{\mathrm d P_z}{\mathrm d t} \sim \beta_{\ell^-, m^-, \ell^+, m^+} \cos(\psi_{\ell^-, m^-, \ell^+, m^+})\,,
\end{equation}
is a combination of the kick amplitude
\begin{equation}
	\beta_{\ell^-, m^-, \ell^+, m^+} = a^-_{\ell^-,m^-} a^+_{\ell^+,m^+} \dot{\phi}^-_{\ell^-,m^-} \dot{\phi}^+_{\ell^+,m^+} \,,
\end{equation}  
and the phase difference
\begin{equation}
	\psi_{\ell^-, m^-, \ell^+, m^+} = \phi^-_{\ell^-,m^-} - \phi^+_{\ell^+,m^+}\,.
\end{equation}

In the vast majority of configurations, the $(\ell^-,m^-,\ell^+,m^+)= (2,2,2,2)$ interactions are the determining factor in the total kick velocity. However, there are two exceptions in which the subdominant interactions could exert a significant influence on the kick. Firstly, the subdominant kick amplitudes are of the same order of magnitude as $\beta_{2,2,2,2}$. Secondly, the cosine of the dominant phase difference at the time of maximal linear momentum emission is close to zero, whereas this is not the case for the subdominant phase differences. 

These two cases are illustrated in Fig.~\ref{fig:subdom_kick}, and are based on two NR simulations of the latest 3.0.0 version of the SXS catalog \cite{SXS_2025_SXSCatalogData300, SXS_2025_SXSPackage}. For \texttt{SXS:BBH:1941} \cite{SXS_2025_SXSBBH1941, Varma_2019_NRSur7dq4}, the kick amplitudes associated with the interactions $(\ell^-,m^-,\ell^+,m^+) = (3,3,3,3)$ and $(\ell^-,m^-,\ell^+,m^+) = (4,4,4,4)$ are of the same order of magnitude to that of the dominant interaction. The effect is so pronounced that neglecting the subdominant multipole asymmetries in the kick calculation would introduce a larger bias than entirely omitting the dominant asymmetry only, as can be seen in the lower left corner of Fig.~\ref{fig:subdom_kick}.

In the case of \texttt{SXS:BBH:3392} \cite{SXS_2025_SXSBBH3392}, the dominant kick amplitude is large. However, since the cosine of the dominant phase difference is close to zero, the dominant interactions contribute negligibly to the final kick value, albeit it has some influence up to $t\approx20\,M$. In contrast, the next-to-leading-order phase differences attain values different from zero, such that their associated kick amplitudes contribute significantly to the final kick. Consequently, neglecting the subdominant asymmetries would result in a bias of approximately $50\,\mathrm{km/s}$ in the final kick.

\subsection{Source characterization with next-generation detectors}
\label{sec:PE}

\begin{table}
	\centering
	\renewcommand{\arraystretch}{1.7}
	\begin{tabular}{cccccc}    
		\toprule
		Parameter & Fig.~\ref{fig:posteriors} & Fig.~\ref{fig:inclination_dependence} (dark) & Fig.~\ref{fig:inclination_dependence} (mid) & Fig.~\ref{fig:inclination_dependence} (light)\\ 
		\midrule
		$M$ & $70\,M_\odot$ & $90\,M_\odot$ & $90\,M_\odot$ & $90\,M_\odot$ \\
		$q$ & $1.5$ & $1$ & $1$ & $1$ \\
		$\vec \chi_1$ & $(0.8, 0, 0)$ & $(0.8, 0, 0)$ & $(0.8, 0, 0)$ & $(0.8, 0, 0)$\\
		$\vec \chi_2$ & $(0.5, 0, 0)$ & $(-0.8, 0, 0)$ & $(-0.8, 0, 0)$ & $(-0.8, 0, 0)$ \\
		$f_{ref}$ & $20$ Hz & $20$ Hz & $20$ Hz & $20$ Hz\\
		$d_L$ & $680$ Mpc & $500$ Mpc & $500$ Mpc & $500$ Mpc\\
		$\iota$ & $\pi/6$ & $\pi/2$ & $\pi/3$ & $0$\\
		\bottomrule
		\hline
	\end{tabular}
	\caption{Injected parameter values for the analyses shown in Fig.~\ref{fig:posteriors} and Fig.~\ref{fig:inclination_dependence}.}
	\label{tab:injection}
\end{table}

\begin{figure*}[tb]
	\includegraphics[width=\linewidth]{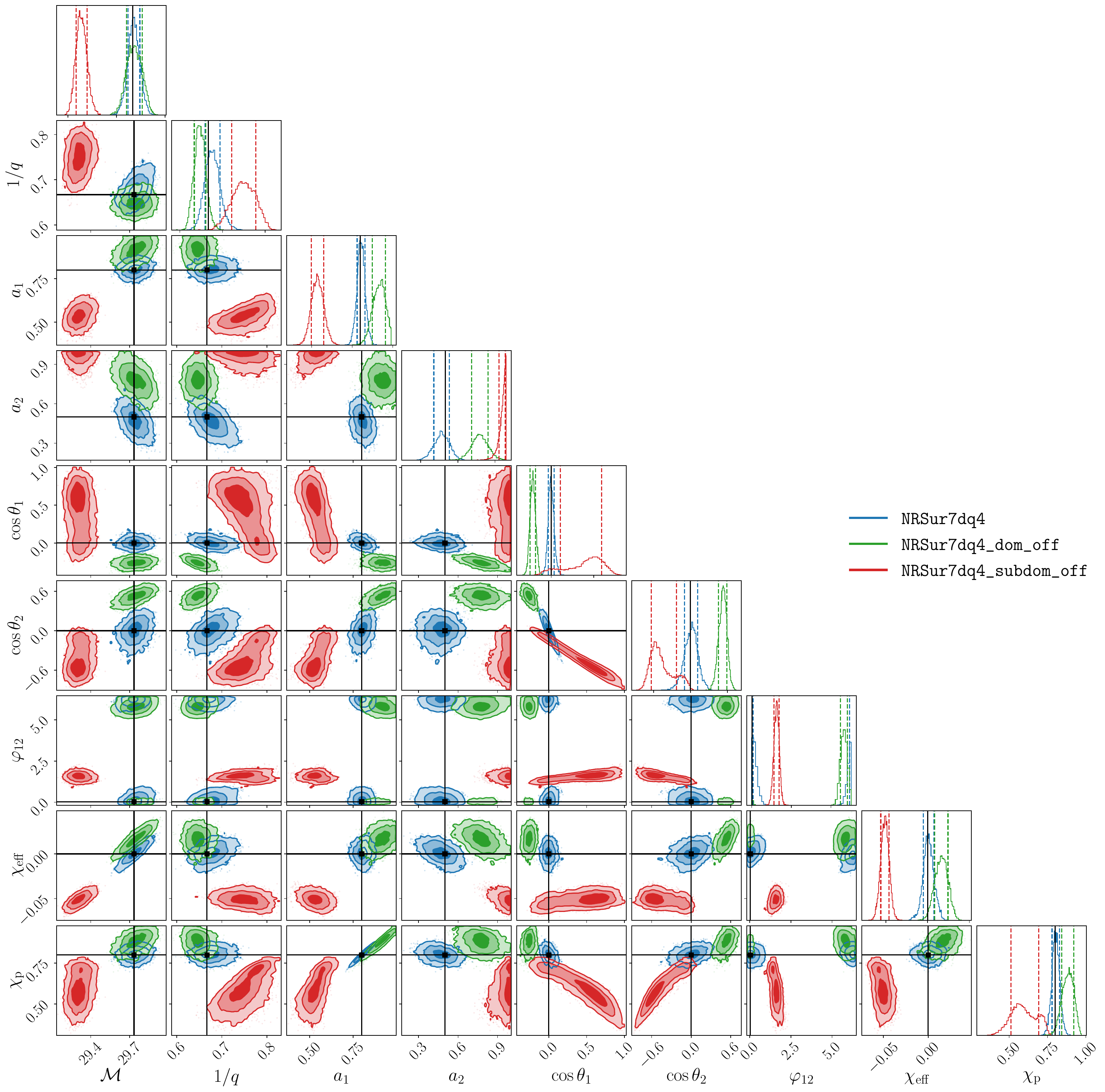}
	\caption{Recovered posteriors of the chirp mass $\mathcal{M}$, the inverted mass ratio $1/q$, the spin magnitudes $a_1$ and $a_2$, the spin tilt angles $\cos\theta_1$ and $\cos\theta_2$, the spin azimuthal angle $\varphi_{12}$, the effective spin $\chi_{\rm eff}$ and the effective spin precession parameter $\chi_{\rm p}$. We use waveform models including all asymmetries, \texttt{NRSur7dq4}, neglecting dominant asymmetries, \texttt{NRSur7dq4\_dom\_off} and neglecting subdominant asymmetries \texttt{NRSur7dq4\_subdom\_off}. Black lines shows the injected values of a \texttt{NRSur7dq4} waveform into the ET-D design sensitivity curve.}
	\label{fig:posteriors}
\end{figure*} 

In addition to the kick velocity, we investigate the role of subdominant multipole asymmetries in estimating the source properties with next-generation ground-based detectors. Subdominant multipole asymmetries contribute minimally to the signal strength. Hence, they will only become relevant for signals with high signal-to-noise ratios. In this section, we investigate how subdominant asymmetries influence parameter estimation with a next-generation detector such as the Einstein Telescope (ET) \cite{Punturo_2010_ET, Hild_2011_ETD, ET_2025_ETScience} based on selected examples.

To infer the source properties we use Bayesian inference. This statistical framework estimates a probability distribution for each parameter using Bayes' theorem, defined as
\begin{equation}
	p(\vec{\theta}|d, h_m) = \frac{\pi(\vec{\theta}|h_m) \mathcal{L}(d|\vec{\theta}, h_m)}{p(d|h_m)}\, .
\end{equation}
As shown by the equation, the posterior probability of a binary signal with parameters $\vec{\theta}$ given the data $d$ and a signal model $h_m$ is proportional to the the prior probability distribution, $\pi(\vec{\theta}|h_m)$, and the likelihood of the data given the parameters $\vec{\theta}$ and a signal model, $\mathcal{L}(d|\vec{\theta},h_m)$. The denominator $p(d|h_m)$ is known as the evidence and is given by the likelihood marginalized over all parameters. In GW data analysis, it is common to assume stationary Gaussian noise. Under this assumption, the likelihood is proportional to 
\begin{equation}
	\mathcal{L}(d|\vec{\theta},h_m) \propto \exp \left(- \braket{d - h_m(\vec{\theta})|d-h_m(\vec{\theta})}/2 \right) \, ,
\end{equation}
where we have introduced the noise-weighted inner product, defined as
\begin{equation}
	\braket{d|h} = 4 \Re \int_{0}^{\infty} \frac{\tilde{d}(f) \tilde{h}^*(f)}{S_n(f)}\, {\rm d}f\, .
\end{equation}
$\tilde{d}$ and $\tilde{h}$ denote the Fourier transforms of $d(t)$ and $h(t)$, $*$ indicates complex conjugation and $S_n$ is the one-sided power spectral density of the detector noise.

In practice, we generate waveforms with limited length. We use $f_{\rm min} =20$ Hz and $f_{\rm max} = 2048$ Hz as lower and upper frequency limits respectively. Assuming the ET-D design power spectral density \cite{Hild_2011_ETD}, we perform injections into zero noise, which effectively corresponds to averaging over many noise realizations, in order to isolate waveform modeling biases from statistical noise effects. To inject signals and infer the marginalized posterior distributions, we use \texttt{Bilby} \cite{Ashton_2019_Bilby} in combination with the nested sampling algorithm \texttt{DYNESTY} \cite{Speagle_2020_Dynesty} using $n_{\text{live}} = 2000$ live points. Posterior samples are post-processed using \texttt{PESummary} \cite{Hoy_2021_PESummary}. We adopt uniform priors on the component masses, $m_i \in [1,1000]\,M_\odot$. The spin magnitudes are drawn from a uniform distribution, $|\vec{\chi}_i| \in [0,0.99]$, while the spin directions are assumed to be isotropically distributed. The luminosity distance prior is taken to be uniform in the range $d_L \in [100,5000]\,\mathrm{Mpc}$.

In our injection and recovery study, we simulate signals from BBH mergers. Tab.~\ref{tab:injection} includes all the relevant parameters of our waveform injection with \texttt{NRSur7dq4}. This system exhibits significant dominant and subdominant multipole asymmetries. This enables us to investigate their influence by recovering with the same waveform model, \texttt{NRSur7dq4}, which incorporates all asymmetries, alongside two artificial waveform models, \texttt{NRSur7dq4\_dom\_off} and \texttt{NRSur7dq4\_subdom\_off}, which can switch off the dominant or subdominant asymmetry, respectively. Note that these models differ from the model \texttt{NRSur7dq4\_sym} used in Ref.~\cite{Kolitsidou_2024_AsymSpinMeasure}, in which all multipole asymmetries are set to zero.

Fig.~\ref{fig:posteriors} shows the posterior distributions for several intrinsic parameters described in the caption of that figure. For this particular injection, we visually identify that the bias of neglecting subdominant asymmetries is equal to or slightly larger than the bias of neglecting only dominant asymmetries, across all parameters. This behavior is not expected to be generic for all signals. However, the present example demonstrates that such a situation can indeed occur.   

Neglecting subdominant asymmetries leads to comparatively small changes in the waveform and therefore to small mismatches. Standard mismatch-based indistinguishability criteria are known to be conservative in this regime and tend to predict parameter biases at lower signal-to-noise ratios (SNR) than actually required \cite{Thompson_2025_indist}, motivating a direct analysis via PE. Here we are in a relatively high SNR regime of around $\rho \approx 220$. Since the systematic error is independent of the SNR while the statistical error scales as $1/\rho$, the qualitative scaling between statistical and systematic uncertainties remains valid. Consequently, this example serves as an illustration of how, at sufficiently high SNR, waveform systematics induced by missing dominant and subdominant asymmetries can affect parameter estimation. However, because the accuracy requirements typically quoted for NR simulations, and therefore for the surrogate models derived from them, are formally exceeded in this regime \cite{Scheel_2025_SXS3rdCatalog, Varma_2019_NRSur7dq4}, the absolute numerical value of the inferred systematical biases should not be over-interpreted. This highlights the need for continued improvements in NR simulations to reliably model high-SNR signals.

\begin{figure}[tb]
	\includegraphics[width=\linewidth]{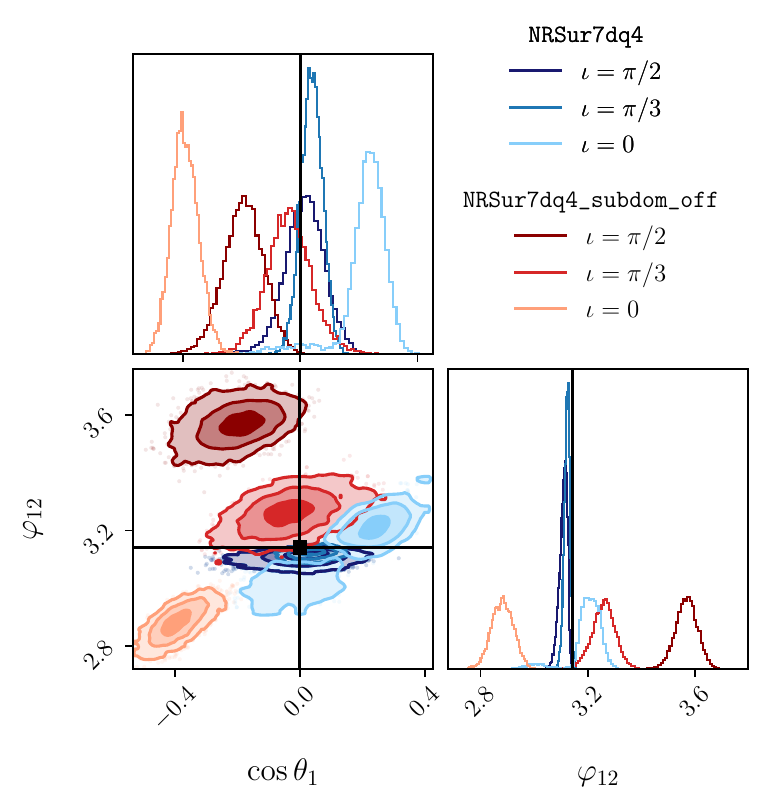}
	\caption{Inclination dependence of the systematic bias between including all asymmetries, \texttt{NRSur7dq4} (blue), and neglecting the subdominant ones, \texttt{NRSur7dq4\_subdom\_off} (red). Because the SWSH have a complex structure, it is non-trivial to predict a priori which inclination leads to the largest bias for a given parameter, here exemplified with the spin tilt angle $\cos \theta_1$ and the spin azimuthal angle $\phi_{12}$. Dark, mid, and light colors indicate inclinations of $\iota = \pi/2$, $\pi/3$, and $0$, respectively.}
	\label{fig:inclination_dependence}
\end{figure} 

In Fig.~\ref{fig:inclination_dependence} we analyze the inclination dependence of the bias due to neglecting subdominant asymmetries. We inject a superkick case with parameters given in Tab.~\ref{tab:injection}. We decide on a superkick system with equal masses and opposite spins aligned in the orbital plane because it exhibits large asymmetries despite not showing any precession in the orbital plane. We use three different inclinations, edge-on $\iota=\pi/2$, face-on $\iota=0$, and an intermediate case $\iota=\pi/3$. The systematic differences between \texttt{NRSur7dq4} and \texttt{NRSur7dq4\_subdom\_off} vary with the inclination in a non-trivial way, reflecting the complex structure of the SWSH visualized in Fig.~\ref{fig:swsh}. The magnitude and sign of the bias depend on the specific parameter. For example, the bias in the spin tilt angle $\cos \theta_1$ is largest for the face-on configuration, whereas for the spin azimuthal angle $\varphi_{12}$ it is maximal for the edge-on configuration. This highlights the need for further investigation to fully quantify and understand the impact of subdominant asymmetries across the parameter space.

\section{Phenomenology of subdominant multipole asymmetries}
\label{sec:phenom}
This section develops phenomenological insights into the physics of subdominant multipole asymmetries. We use NR waveforms, PN predictions and analytical QNM models to derive formulas and propose ideas for potential future models of subdominant multipole asymmetries. We base our work on the existing time-domain EOB model for $\ell=m$ asymmetries \cite{Estelles_2025_EOBasym} and analyze separately the inspiral-plunge and merger-ringdown regime for all subdominant asymmetries.

\subsection{Inspiral-plunge}
\label{sec:inspiral}

\begin{figure*}[tb]
	\includegraphics[width=\linewidth]{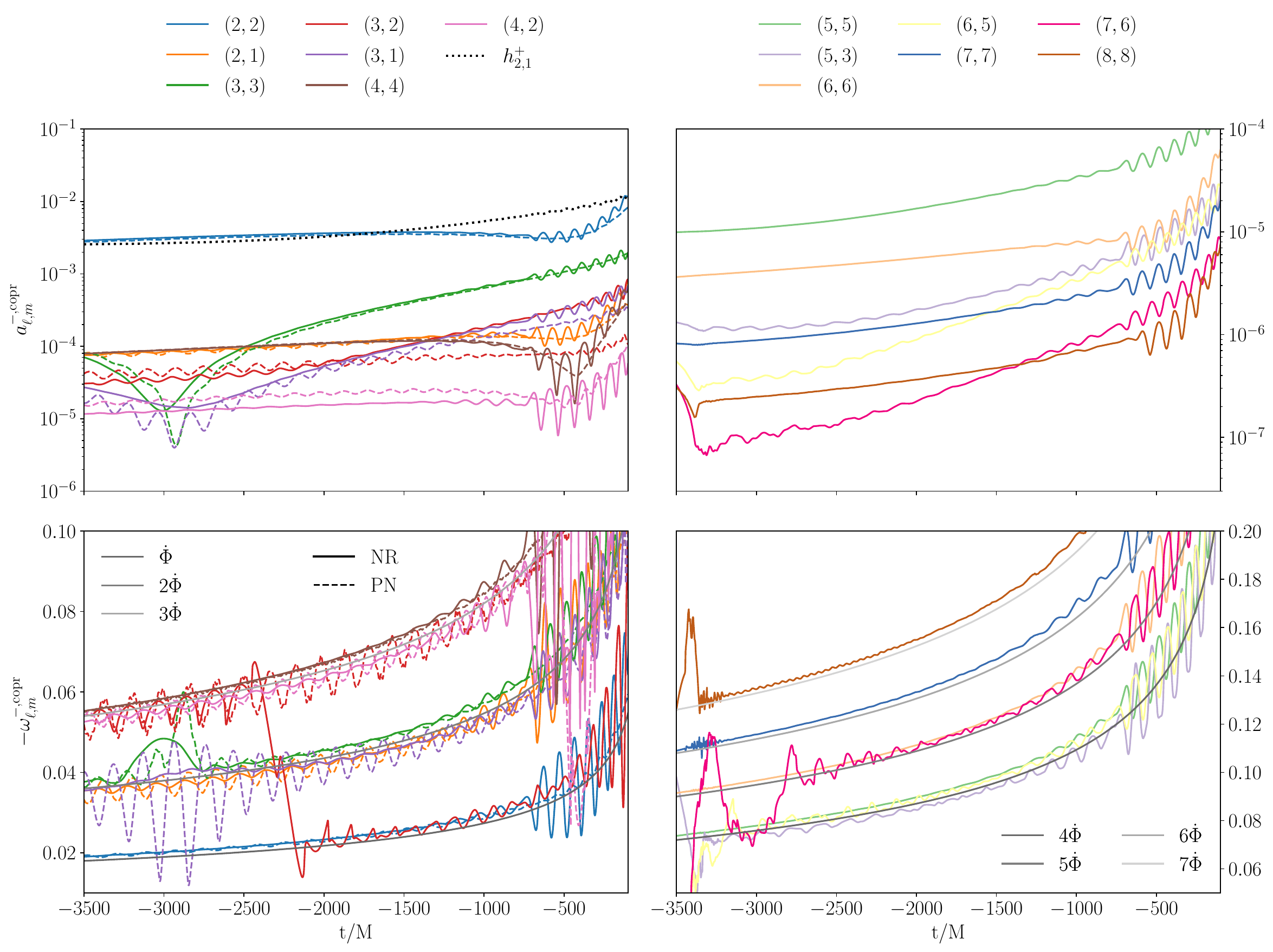}
	\caption{Antisymmetric amplitudes and frequencies in the inspiral regime, exemplified by \texttt{SXS:BBH:1207}, a $q=2$ system with large spin components in the orbital plane. The left panel displays multipoles for which PN results are available, while the right panel shows higher-order antisymmetric multipoles. For reference, the symmetric $(2,1)$ amplitude is included as a dotted black line to illustrate the relative order of magnitude. The antisymmetric frequencies scale as $m \pm 1$ times the orbital frequency $\dot\Phi$.}
	\label{fig:inspiral_asym_wf}
\end{figure*}

\begin{figure}[tb]
	\includegraphics[width=.66\linewidth]{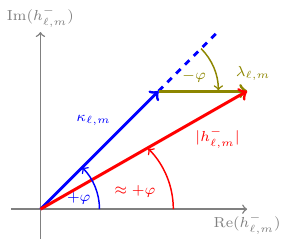}
	\caption{Illustration of Eq.~\eqref{eq:antisym_PN_kl_form} for the case $\kappa_{\ell,m}, \lambda_{\ell,m} \in \mathbb{R}^+$ with $\kappa_{\ell,m} > \lambda_{\ell,m}$. The resulting sum $h^-_{\ell,m}$ exhibits an average phase rotation in the positive $\varphi$ direction.}
	\label{fig:illustration_kl_PN_argument}
\end{figure} 

The antisymmetric components of the waveform only grow significantly in the merger regime. Consequently, in the inspiral regime, the contributions from antisymmetric multipoles are suppressed by several orders of magnitude compared to the symmetric contributions. Nevertheless, the underlying physics remains of considerable interest in this regime. 

In Fig.~\ref{fig:inspiral_asym_wf}, we illustrate the characteristics of subdominant and dominant antisymmetric amplitudes, and their associated frequencies in the coprecessing frame for \text{SXS:BBH:1207} \cite{SXS_2025_SXSBBH1207}. 

The amplitudes show good agreement with PN predictions \cite{Boyle_2014_PrecessingBHOperators} implemented in \cite{Boyle_2024_PNjulia}, which are available for the $(2,2), (2,1), (3,3), (3,2), (3,1), (4,4), (4,2)$ antisymmetric multipoles. Minor discrepancies arise, particularly in resolving the small oscillations. However, on average the results show quantitative agreement and remain in the same order of magnitude throughout the inspiral. In Tab.~\ref{tab:antisym_PN_kl_q}, the amplitudes are summarized for the special case of single in-plane-spin configurations. This choice is motivated by the fact that most of the antisymmetric amplitude contribution is expected to arise from the mass ratio and in-plane-spin magnitude.

For the $(3, 2)$ antisymmetric NR amplitude and frequency, we found larger differences to PN in several cases, which will be discussed at the end of this section. In addition to this multipole, Ref.~\cite{Estelles_2025_EOBasym} also reported significant discrepancies for the $(2, 1)$ multipole. However, we did not observe this in most of the cases, as the implementation of the antisymmetric PN multipoles \cite{Boyle_2024_PNjulia}, which we are using, already accounts for 2PN spin-spin term corrections reported in the documentation. These corrections were not used by Ref.~\cite{Estelles_2025_EOBasym}.

In the following, our analysis will focus on the frequencies of the antisymmetric multipoles. We observed a universal characteristic that also holds for higher multipoles that are not currently accessible within the PN framework. On average, the antisymmetric frequencies follow integer multiples of the orbital frequency $\dot{\Phi}$, as initially already reported by Boyle et. al \cite{Boyle_2014_PrecessingBHOperators}. In the coprecessing frame it is $m\pm1$ times $\dot{\Phi}$, as can be seen from the example in Fig.~\ref{fig:inspiral_asym_wf}. Note that for $m=0$ antisymmetric multipoles, the frequency is zero by definition. If the metadata of the NR simulations did not provide it, we used the orbital phase derived via \cite{Varma_2019_NRSur7dq4}
\begin{equation}
	\Phi = \frac{1}{4} \left[\arg\left(h_{2,-2}^\text{copr}\right) - \arg\left(h_{2,2}^\text{copr} \right)\right] \,.
\end{equation}

\begin{table*}
	\centering
	\renewcommand{\arraystretch}{2.75}
	\begin{tabular}{cccc}    
		\toprule
		$(\ell,m)$ & $\kappa_{\ell,m}$ & $\lambda_{\ell,m}$ & $\mathfrak{m}^-_{\ell,m}$\\
		\midrule
		$(2,1)$ & $\displaystyle \frac{(17 + 12 q) q \chi v^3}{12 (1 + q)^2}$ & $\displaystyle \frac{(3 + 4 q) q\chi v^3}{4 (1 + q)^2}$ & $m+1=2$ \\
		$(2,2)$ & $0$ & $\displaystyle\frac{i q \chi v^2}{2(1+q)}$ & $m-1=1$ \\
		$(3,1)$ & $\displaystyle\frac{\sqrt{2}\,q\chi v^3}{3\sqrt{7}(1+q)^2}$ & $0$ & $m+1=2$ \\
		$(3,2)$ & $\displaystyle -\frac{i \sqrt{5} (-69 + 40 q + 56 q^2) q \chi v^4}{48 \sqrt{7} (1 + q)^3}$ & $\displaystyle -\frac{i \sqrt{5} (-35 - 8 q + 40 q^2) q\chi v^4}{48 \sqrt{7}(1 + q)^3}$ & $m+1=3$ \\ 
		$(3,3)$ & $0$ & $\displaystyle-\frac{\sqrt{10}q\chi v^3}{ \sqrt{21}(1+q)^2}$ & $m-1=2$ \\
		$(4,2)$ & $\displaystyle \frac{ i 9 \sqrt{5} q \chi v^4 }{112 (1 + q)^3}$ & $\displaystyle \frac{i \sqrt{5} q \chi v^4}{336 (1 + q)^3}$ & $m+1=3$ \\
		$(4,4)$ & $0$ & $\displaystyle -\frac{ i9 \sqrt{5} q \chi v^4 }{8\sqrt{7} (1 + q)^3}$ & $m-1=3$ \\
		\bottomrule
		\hline
	\end{tabular}
	\caption{Amplitudes $\kappa_{\ell,m}$ and $\lambda_{\ell,m}$ according to Eq.~\eqref{eq:antisym_PN_kl_form} for single-in-plane-spin configurations, obtained using the PN expressions in App.~A of Ref.~\cite{Boyle_2014_PrecessingBHOperators}. If $\kappa_{\ell,m}$ ($\lambda_{\ell,m}$) dominates, the coprecessing frequency ratio is $\mathfrak{m}^-_{\ell,m} = m+1$ ($\mathfrak{m}^-_{\ell,m} = m-1$).}
	\label{tab:antisym_PN_kl_q}
\end{table*}

\begin{figure}[tb]
	\includegraphics[width=\linewidth]{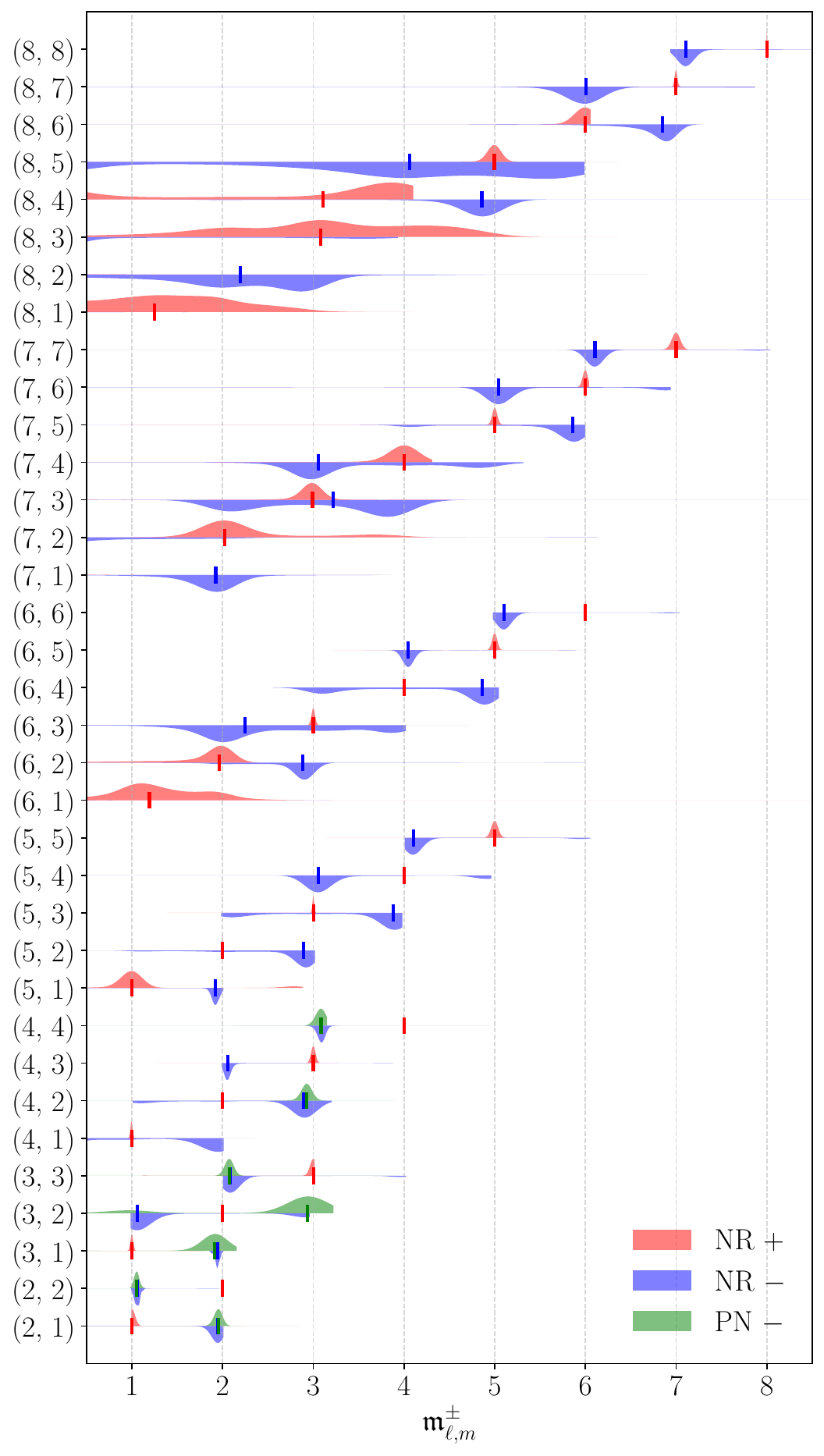}
	\caption{Frequency ratio distribution across all available multipoles calculated with all quasi-circular precessing simulations of the third \texttt{SXS} catalog. The symmetric frequency ratios in red follow the well-known relation $\mathfrak{m}^+_{m,\ell}=m$, while the antisymmetric frequency ratios in blue behave as $\mathfrak{m}^-_{\ell,m} = m \pm 1$. The antisymmetric PN frequency ratio distributions, which have the same intrinsic parameters as those in the NR catalog, are shown in green. The median values of the distributions are indicated by vertical lines.}
	\label{fig:mf_pm_violin_plot_tref_tm100_copr}
\end{figure}

To understand when it is $m+1$ or $m-1$, we consider the PN expressions given in App.~A of Ref.~\cite{Boyle_2014_PrecessingBHOperators}. There, the second bracket terms of the Eqs.~(A5) correspond to the antisymmetric waveform in the corotating frame. 

In order to gain insight into the structure of antisymmetric PN frequencies, the we make the following considerations. We only use the leading order expressions and assume a binary with the heavier BH spinning entirely in the orbital plane. Then we can write the symmetric and antisymmetric spins defined in Ref.~\cite{Boyle_2014_PrecessingBHOperators} as
\begin{subequations}
\label{eq:antisym_sym_spin}
\begin{eqnarray}
	S_n &= \frac{q^2M^2}{(1+q)^2}\,\chi \cos(\varphi)\,,  \\
	S_\lambda &= \frac{q^2M^2}{(1+q)^2}\,\chi \sin(\varphi)\,,  \\
	\Sigma_n &= -\frac{qM^2}{1+q}\,\chi \cos(\varphi) \,, \\
	\Sigma_\lambda &= -\frac{qM^2}{1+q}\,\chi \sin(\varphi)\,.
\end{eqnarray}
\end{subequations}

Here $\varphi(t)$ is the in-plane spin angle in the corotating frame and $\chi$ the spin magnitude. By inserting Eqs.~\eqref{eq:antisym_sym_spin} in Eqs.~(A5) of Ref.~\cite{Boyle_2014_PrecessingBHOperators} and using Euler's formula, we can derive a general form of all antisymmetric multipoles,
\begin{equation}
\label{eq:antisym_PN_kl_form}
	h_{\ell,m}^{-, \mathrm{coro}} = \kappa_{\ell,m}\, e^{+i\varphi} + \lambda_{\ell,m}\, e^{-i\varphi}\,,
\end{equation}
where $(\kappa_{\ell,m}, \lambda_{\ell,m}) \in (\mathbb{R} \times \mathbb{R}) \cup (i\mathbb{R} \times i\mathbb{R})$ are either both purely real or both purely imaginary functions. This means that on average the phase in Eq.~\eqref{eq:antisym_PN_kl_form} rotates with $+\varphi(t)$ if $\left|\kappa_{\ell,m}\right| > \left|\lambda_{\ell,m}\right|$ and with $-\varphi(t)$ if $\left|\kappa_{\ell,m}\right| < \left|\lambda_{\ell,m}\right|$, see the illustration in Fig.~\ref{fig:illustration_kl_PN_argument}. Note that we only have a constant phase shift of $\pi/2$ when both $\kappa_{\ell,m}, \lambda_{\ell,m}$ are imaginary. In Tab.~\ref{tab:antisym_PN_kl_q} we summarize our findings for all available antisymmetric PN multipoles. 

Corotating and coprecessing frame multipoles are related via 
\begin{equation}
	h_{\ell,m}^{\text{copr}} = h_{\ell,m}^{\text{coro}} \cdot e^{-im\Phi} \,.  
\end{equation}
The azimuthal angle in the corotating frame can be expressed via (cf. paragraph above Eq.~(A3) in Ref.~\cite{Boyle_2014_PrecessingBHOperators})
\begin{equation}
	\varphi = \frac{\pi}{2} - \Phi - \alpha\,,
\end{equation}
where $\alpha(t)$ is the precession angle in the coprecessing frame. Thus, in agreement with Ref.~\cite{Estelles_2025_EOBasym}, we can write the coprecessing frequency as
\begin{equation}
\label{eq:antisym_copr_freq}
	\omega^{-,\text{copr}}_{\ell,m} \approx \pm \dot{\varphi} - m\dot{\Phi} = \mp \dot\alpha - (m \pm 1)\dot{\Phi} \,.
\end{equation}
Note that the sign is dependent on the dominance of either $\kappa_{\ell,m}$ ($m+1$) or $\lambda_{\ell,m}$ ($m-1$). Assuming that the precession frequency $\dot{\alpha}$ is small compared to the orbital frequency $\dot{\Phi}$, we can give a simple approximation of the antisymmetric frequency ratio 
\begin{equation}
\label{eq:antisym_freq_ratio}
	\mathfrak{m}^-_{\ell,m} = \frac{\omega^{-,\text{copr}}_{\ell,m}}{-\dot{\Phi}} \approx m \pm 1 \,,
\end{equation} 
as phenomenologically already observed for NR and PN waveforms in Fig.~\ref{fig:inspiral_asym_wf}. 

The simple relation is an universal feature across all multipoles and only weakly depend on the parameter space. To show this, we performed a systematic study of all non-eccentric and precessing NR waveforms in the SXS catalog. We computed the ratio of the coprecessing antisymmetric/symmetric frequencies and $-\dot{\Phi}$. Then we extracted the median value of this ratio in a window ranging from the reference time to up to 100 M before merger. The distributions across all simulations are shown in Fig.~\ref{fig:mf_pm_violin_plot_tref_tm100_copr}.

The well-known symmetric frequency ratio $\mathfrak{m}^+_{\ell,m} = m$ is correctly recovered by the algorithm. An exception arises for very small subdominant multipoles, such as the (8,2) or (7,2) multipoles, where numerical noise could obscure the extraction of a clean frequency ratio.

Nevertheless, the algorithm can be considered reliable for inferring both symmetric and antisymmetric frequency ratios. This is particularly true given that the PN frequency ratios are consistently recovered, except in the case of the $(3,2)$ ratio. The NR distributions for $m = \ell$ and $m = \ell - 1$ peak at $m - 1$. The $(2,1)$ ratio peaks at $m+1$, because $m-1$ would otherwise correspond to a zero frequency. For all other values of $m$, the antisymmetric ratios are mostly at $m + 1$.

Another observation concerns the slight offset of the median values of the distribution from the exact positions at $m \pm 1$. Specifically, the median for $m+1$ is shifted towards lower values, while the median for $m-1$ is shifted towards higher values. This shift arises from the sign of the precession frequency $\dot{\alpha}$, which, according to Eq.~\eqref{eq:antisym_copr_freq}, is opposite to the sign associated with $m \pm 1$. 

We do not obtain consistent results for the $(3, 2)$ antisymmetric multipole. The NR frequency occasionally exhibits jumps between one and three times the orbital frequency. The PN frequency remains reliably at three times the orbital phase up to mass ratios of approximately 3.5. For larger mass ratios, however, the frequency is dominated by numerical noise. We examined several hypotheses, such as the possibility that $a^-_{3,2}$ is very small or that, for different mass ratios, the magnitudes of the parameters $\kappa_{3,2}$ and $\lambda_{3,2}$ alternate. None of these provided a satisfactory explanation. Therefore, we suggest that the $(3, 2)$ multipole should be treated with caution.

\subsection{Merger-ringdown}
\label{sec:merger-ringdown}

\begin{figure*}[tb]
	\includegraphics[width=\linewidth]{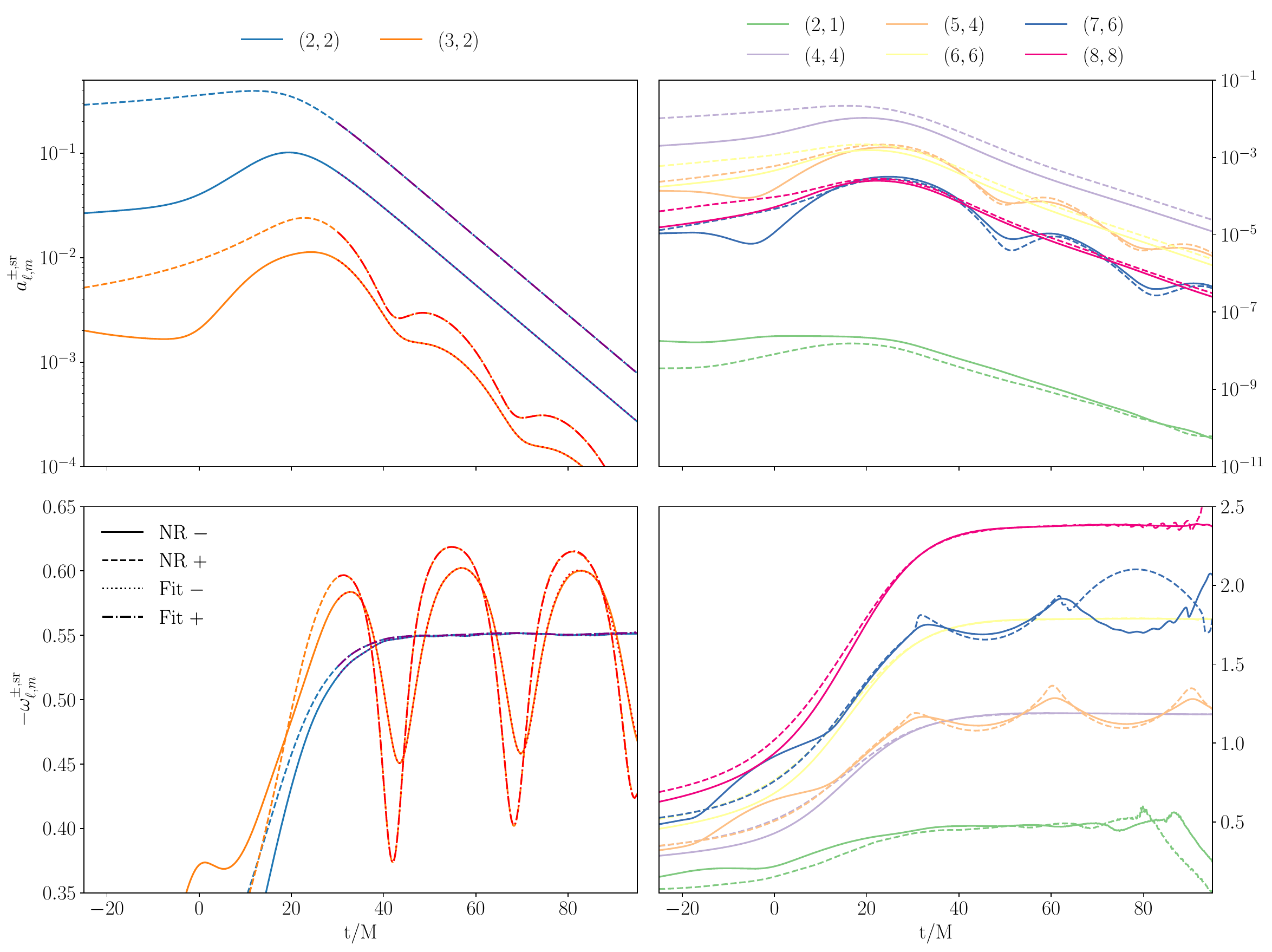}
	\caption{Super rest frame antisymmetric and symmetric amplitudes and frequencies in the merger-ringdown regime, shown for \texttt{SXS:BBH\_ExtCCE:0009}, a $q=1$ superkick configuration. The average symmetric and antisymmetric frequencies coincide. The left panel compares the fit model of Eq.~\eqref{eq:QNM_analytical} with the NR data, showing excellent agreement.}
	\label{fig:ringdown_asym_wf}
\end{figure*} 

The dominant and subdominant antisymmetric amplitude are maximal in the merger regime. As we saw in Sec.~\ref{sec:kick}, the phase difference between the antisymmetric and symmetric waveforms at merger is crucial for the final kick. Unfortunately, we lack an analytical description of this regime, so we are always dependent on calibrating to NR simulations to accurately describe the amplitudes and phases. 

However, the radiation emitted by the remnant BH during its ringdown phase is very well understood by perturbation theory. The most general spherical multipole-decomposed analytical QNM model in the super rest frame is given by \cite{Zertuche_2022_BMSFrameQNM}
\begin{equation}
\label{eq:QNM_analytical}
\begin{aligned}
	h^{\text{sr}}_{\ell,m} (t) 
	= \sum_{\ell', n, p} \mathcal{A}_{\ell,m,n,p} \, &e^{-i\omega_{\ell',m,n,p} \left(t-t_0\right)} \\
	&\cdot C_{\ell,\ell',m}\left(a\omega_{\ell',m,n,p}\right)\,, 
\end{aligned}
\end{equation}
where $n=0,1,...$ is the overtone index, $\mathcal{A}_{\ell,m,n,p}$ are the QNM amplitudes, $\omega_{\ell,m,n,p}$ are the complex QNM frequencies, $t_0$ is the starting time of the ringdown, $C_{\ell,\ell',m}$ are the spherical-spheroidal mixing coefficients and $a$ is the Kerr parameter of the final BH. Ordinary modes have $p=\text{sgn}\left[\text{Re}\left(\omega\right)\right] = +1$, while mirror modes have $p=-1$.

In the non-precessing case where Eq.~\eqref{eq:mode_symmetry} is valid, the symmetry relations 
\begin{subequations}
\label{eq:qnm_symmetries}
\begin{eqnarray}
	\omega_{\ell,m,n,-p} &=& -\omega^*_{\ell,-m,n,p} \,,\\
	C^*_{\ell,\ell',-m}(a\omega_{\ell', -m,n,p}) &=& (-1)^{-\ell-\ell'} \\
	&\quad& \cdot\, C_{\ell,\ell',m}(a\omega_{\ell',m,n, -p}) \notag
\end{eqnarray}
\end{subequations}
lead to a simple relation for the QNM amplitudes, 
\begin{equation}
	\mathcal{A}_{\ell,m,n,p} = (-1)^{\ell}\mathcal{A}^*_{\ell,-m,n,-p} \,.
\end{equation}

However, this ringdown amplitude symmetry is broken in the precessing case, while the QNM frequency and mixing coefficient symmetry is preserved. This means that, in the super rest frame, the symmetric and antisymmetric waveform definitions in Eq.~\eqref{eq:plus_minus_wf_def} lead to the following form in the ringdown regime:
\begin{equation}
\label{eq:plus_minus_wf_rd}
\begin{aligned}
	h^{\text{sr},\pm}_{\ell,m} = &\sum_{\ell', n, p} \mathcal{A}^\pm_{\ell',m,n,p} \\
	&\cdot\,e^{-i\omega_{\ell',m,n,p}\left(t-t_0\right)} \,C_{\ell,\ell',m}\left(a\omega_{\ell',m,n,p}\right) \,. 
\end{aligned}
\end{equation}
Here, we defined the antisymmetric ($-$) and symmetric ($+$) ringdown amplitudes
\begin{equation}
	\mathcal{A}^\pm_{\ell,m,n,p} = \frac{1}{2} \left[\mathcal{A}_{\ell,m,n,p} \pm (-1)^{\ell} \mathcal{A}^*_{\ell,-m,n,-p} \right] \,.
\end{equation}

Since the QNM frequencies and mixing coefficients are known from perturbation theory and depend only on the remnant properties, the key to model multipole asymmetries in the ringdown regime lies in accurately capturing the $+/-$ ringdown amplitudes. These amplitudes represent excitations, which depend on the intrinsic parameters of the binary prior to merger. Consequently, fitting to NR is necessary. Since Eq.~\eqref{eq:plus_minus_wf_rd} is analogous to Eq.~\eqref{eq:QNM_analytical}, the basis for fitting procedures, existing QNM fitting algorithms can be applied directly to the $+/-$ waveform in the ringdown regime, enabling a consistent and systematic modeling of these waveforms.

This amplitude fitting has, for instance, been performed in Ref.~\cite{Estelles_2025_EOBasym} for the dominant antisymmetric multipole. In Eq.~(61) there, we can find analogies to our Eq.~\eqref{eq:plus_minus_wf_rd} when neglecting mode-mixing, i.e.~$C_{\ell,\ell',m}=\delta_{\ell\ell'}$, mirror modes, and when only considering the fundamental $n=0$ mode. Even though this equation is used to connect to the inspiral waveform, the key fitting quantities $c_{i,f}^{22,\text{asym}}$ effectively encode what we generally write as $\mathcal{A}^-_{2,2,0,1}$. For the other antisymmetric multipoles with $\ell=m$, the authors of Ref.~\cite{Estelles_2025_EOBasym} found a simple relationship with the symmetric fitting amplitudes, meaning they did not have to fit the antisymmetric case.

The magnitude of the QNM amplitudes is important for determining the average ringdown frequency of a multipole. In Ref.~\cite{Hamilton_2023_RDfreq}, using an argument analogous to the one we used in the inspiral regime via the dominance of amplitudes in Eq.~\eqref{eq:antisym_PN_kl_form}, it was shown that the average ringdown frequency corresponds to the $p=1$ QNM frequency if the associated QNM amplitude dominates over that with $p=-1$, and vice versa. 

This argument can also be applied to Eq.~\eqref{eq:plus_minus_wf_rd}. The norm of the $+/-$ ringdown amplitudes is  
\begin{equation}
\begin{aligned}
	\left|\mathcal{A}^\pm_{\ell,m,n,p}\right|^2 &= \frac{1}{2}\left|\mathcal{A}_{\ell,m,n,p}\right|^2 + \frac{1}{2} \left|\mathcal{A}_{\ell,-m,n,-p}\right|^2 \\
	&\quad \pm (-1)^{\ell} \, \text{Re}\left(\mathcal{A}_{\ell,m,n,p}\,\mathcal{A}_{\ell,-m,n,-p}\right) \,.
\end{aligned}
\end{equation}
Simple complex algebra shows that the last term is always smaller than, or equal to, the two preceding terms. Therefore, the mean value of the ringdown frequency can be approximated by $\omega_{\ell,m,n,+1}$ if 
\begin{equation}
\label{eq:square_ampl}
\begin{aligned}
\left|\mathcal{A}_{\ell,m,n,+1}\right|^2 &+ \left|\mathcal{A}_{\ell,-m,n,-1}\right|^2 \\
&>\left|\mathcal{A}_{\ell,m,n,-1}\right|^2 + \left|\mathcal{A}_{\ell,-m,n,+1}\right|^2,
\end{aligned}
\end{equation}
and by $\omega_{\ell,m,n,-1}$ in the opposite case. In particular, it also means that antisymmetric and symmetric multipoles share the same average frequency. 

We have observed this average frequency characteristic for all sufficiently strong subdominant multipole asymmetries in the super rest frame. An illustrative example for a superkick case is shown in Fig.~\ref{fig:ringdown_asym_wf}. To generate this plot, we employed the \texttt{qnmfits} package of the SXS Collaboration \cite{Zertuche_2025_qnmfits}. In this framework, the QNM frequencies and spherical–spheroidal mixing coefficients are provided through the \texttt{qnm} package \cite{Stein_2019_qnm}, which implements the Cook–Zalutskiy spectral method \cite{Cook_2014_qnmfreqs}. A ringdown starting time of $t_0=30\,M$ and including mode mixing of the $(2,\pm2)$ and $(3,\pm2)$ multipoles provided an accurate fit to the $+/–$ waveform in the ringdown regime.

The fit also reproduces the small oscillations around the average frequency. These arise from the superposition with the non-dominant $p$ contribution in Eq.~\eqref{eq:plus_minus_wf_rd}. Moreover, the small effect of the final mixing term in Eq.~\eqref{eq:square_ampl} is apparent, as the symmetric and antisymmetric oscillations are not perfectly identical.

In addition, Fig.~\ref{fig:ringdown_asym_wf} shows the usual $+/-$ amplitudes. In general, the symmetric amplitude remains larger than the antisymmetric one throughout the merger-ringdown regime. However, for certain subdominant multipoles, the amplitudes can become comparable or, in some cases, the antisymmetric component may even exceed the symmetric one. This amplitude peak structure plays a crucial role in determining the kick. However, it can only be accurately assessed through NR simulations and will be addressed in future work, particularly to analyze the high-asymmetry region on the right-hand side of the spin–asymmetry–kick correlation plots. 

\begin{figure}[tb]
	\includegraphics[width=\linewidth]{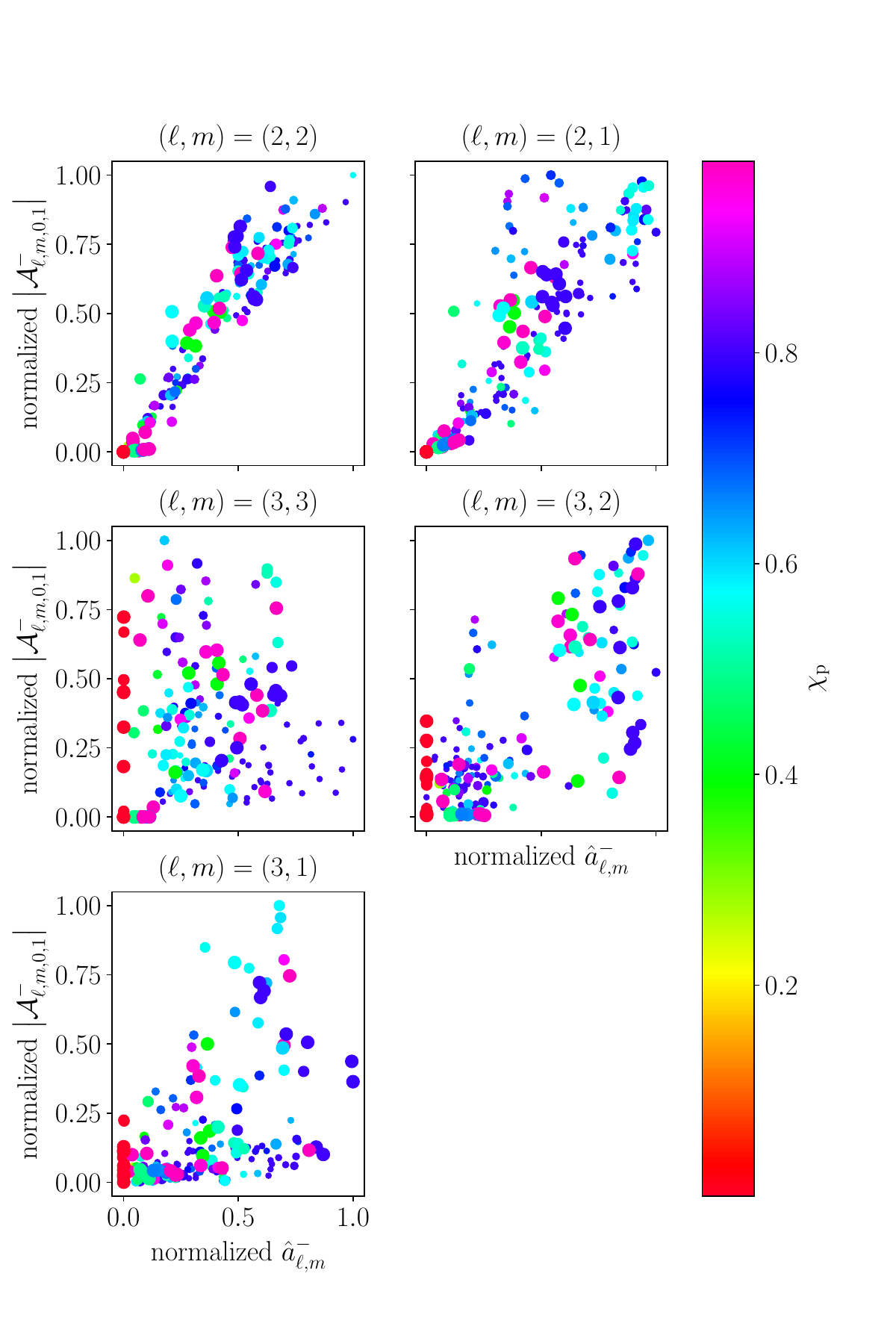}
	\caption{Relation of antisymmetric ringdown amplitudes defined around $t\approx30\,M$ to intrinsic spin parameters of the system and to the peak of the antisymmetric waveform defined around $t\approx10\,M$. The marker size is proportional to the mass ratio.}
	\label{fig:asym_rd_ampls}
\end{figure} 

Nevertheless, the antisymmetric peak amplitudes in the merger regime can be estimated by analyzing the $A^{\pm}_{\ell,m,n,p}$, which are defined shortly after the merger and are therefore highly correlated with the merger peak amplitudes. There exists a broad scientific community dedicated to the development of highly accurate ringdown fitting algorithms \cite{Cook_2020_qnmmultimode, Mitman_2025_VarPro} and novel ringdown models \cite{Pacillio_2024_postmerger, Zertuche_2025_NRSur2dq8RD}. Among these outstanding contributions, we have selected a fitting catalog of 250 waveforms in the super rest frame produced by Zhu et al. \cite{Zhu_2025_BHSpectroscopy, Zhu_2025_git}.

Following the phenomenological approach of Nobili et al. \cite{Nobili_2025_RDAmpl}, who analyzed the dependence of ringdown amplitudes on intrinsic parameters and searched for relations among them, we adopt a similar strategy here. We investigate the correlations between fundamental antisymmetric ringdown amplitudes $\mathcal{A}^-_{\ell,m,0,1}$ around $t\approx 30\,M$ calculated from the aforementioned catalog, the system's intrinsic parameters and the amplitude peaks $\hat{a}^-_{\ell,m}$ around $t\approx10$ in the inertial frame. In Fig.~\ref{fig:asym_rd_ampls}, the color scale encodes $\chi_\text{p}$, while the marker size indicates the mass ratio. Note that the horizontal and vertical axis are normalized to its maximum across all systems for each $(\ell,m)$. 

Several representative correlations can be identified. For all $(\ell,m)$, the amplitudes $\mathcal{A}^-_{\ell,m,0,1}$ show a clear increasing trend with $\chi_\text{p}$. Moreover, for $\ell=2$ there is only a modest difference between the peak amplitudes defined near $t \approx 0$ and the antisymmetric ringdown amplitudes evaluated at $t = 30\,M$. For the $(3,1)$ multipole, equal-mass systems tend to correspond to relatively small antisymmetric ringdown amplitudes but comparatively large peak amplitudes.

Future waveform models may uncover additional correlations between the merger and ringdown regimes, as well as between the inspiral and merger regime. Exploiting these correlations could enable the construction of complete waveforms of subdominant asymmetries, without requiring NR simulations across the entire parameter space.

\section{Conclusions} 
\label{sec:conclusions}
Asymmetries between the subdominant $+m$ and $-m$ multipoles are one of the few remaining unmodeled effects in non-eccentric, precessing waveforms. First, we discussed their importance regarding the kick and source characterization with next-generation ground-based detectors. Second, we identified characteristics in the phenomenology of their waveforms in the inspiral-plunge and merger-ringdown regime, that can be used to develop future models. Here we summarize our main results:
\begin{enumerate}[(i)]
    \item A complete description of the out-of-plane kick requires the inclusion of subdominant asymmetries, as they can make a non-negligible contribution of over 200 km/s to the recoil velocity.
    \item Two mechanisms can lead to a kick with significant subdominant asymmetry contribution: A large subdominant kick amplitude $\beta_{\ell^-, m^-, \ell^+, m^+}$ with $\left(\ell^-, m^-, \ell^+, m^+\right) \neq (2,2,2,2)$, or approximately orthogonal dominant antisymmetric and symmetric waveform contributions at the time of maximal linear momentum emission, so that $\cos(\psi_{2,2,2,2}) \approx 0$.
    \item For the first time, we performed a high-SNR injection-recovery study using a proposed ET detector sensitivity curve to analyze the influence of subdominant asymmetries. We found that, for certain systems, neglecting subdominant asymmetries can introduce a bias in the recovered posterior that is even larger than the bias introduced by neglecting dominant asymmetries. This bias is non-trivially dependent on the inclination of the system.  
    \item In the inspiral regime, the antisymmetric frequencies in the coprecessing frame obey a universal relation to the orbital frequency, $\omega^-_{\ell,m}\sim(m\pm1)\,\dot\Phi$. The sign depends on the specific $(\ell,m)$ multipole. We determined it using both PN and NR results, which are mutually consistent, except for the $(3,2)$ case.
    \item During ringdown, the average $\omega^-_{\ell,m}$ approaches the corresponding symmetric frequency. Depending on combinations of the $+m$ and $-m$ ringdown amplitudes, this frequency can be approximated either by the corresponding ordinary or by the mirror complex frequency. 
    \item Existing QNM fitting algorithms can be used to fit antisymmetric ringdown amplitudes. These amplitudes correlate with the system’s intrinsic parameters, such as a general increasing trend with $\chi_\text{p}$. 
\end{enumerate} 

For further research on subdominant asymmetries, more accurate NR simulations are required. In particular, these would help to resolve the inconsistencies in the antisymmetric frequency ratio of the $(3,2)$ multipole between PN and NR. Including higher-order PN terms as well as additional multipoles in the PN approach would also be beneficial. The dependence of the kick terms on the in-plane spin angles, as well as the fitting of QNM amplitudes, likewise relies on improved NR data. Regarding our PE study, further investigations could be performed by systematically varying intrinsic parameters, since we only considered a limited set of special cases. Moreover, the inclination dependence of the posteriors should be examined by switching individual multipole asymmetries on and off, rather than deactivating the entire set of subdominant asymmetries.

Software - \texttt{NumPy}~\cite{Harris_2020_NumPy}, \texttt{Matplotlib}~\cite{Hunter_2007_Matplotlib}, \texttt{SciPy}~\cite{Virtanen_2020_SciPy}, \texttt{LALSuite}~\cite{LVK_2018_LALSuite}, \texttt{sxs}\cite{SXS_2025_SXSPackage}, \texttt{scri}~\cite{Boyle_2020_scri}, \texttt{GWSurrogate}~\cite{Field_2013_GWSurrogate}, \texttt{PESummary} \cite{Hoy_2021_PESummary}, \texttt{Bilby} \cite{Ashton_2019_Bilby} and \texttt{qnmfits} \cite{Zertuche_2025_qnmfits}.

\section*{Acknowledgments}
We are especially grateful to Shrobana Ghosh for initiating this project and for helpful early discussions. We thank Héctor Estellés Estrella for very useful discussions about the phenomenology of antisymmetric multipoles. We thank Charlie Hoy and Max Melching for their assistance with parameter estimation for next-generation detectors. We thank Jonathan Thompson for comments during the LIGO-Virgo-KAGRA internal review. This work was supported by an Independent Research Group Grant of the Max Planck Society. Computations were performed on the ``Holodeck'' cluster of the Max Planck Independent Research Group ``Binary Merger Observations and Numerical Relativity".\\

\section*{Data availability}
The data supporting this study’s findings are available within the article.

\bibliography{references}

\begin{thebibliography}{85}%
\makeatletter
\providecommand \@ifxundefined [1]{%
 \@ifx{#1\undefined}
}%
\providecommand \@ifnum [1]{%
 \ifnum #1\expandafter \@firstoftwo
 \else \expandafter \@secondoftwo
 \fi
}%
\providecommand \@ifx [1]{%
 \ifx #1\expandafter \@firstoftwo
 \else \expandafter \@secondoftwo
 \fi
}%
\providecommand \natexlab [1]{#1}%
\providecommand \enquote  [1]{``#1''}%
\providecommand \bibnamefont  [1]{#1}%
\providecommand \bibfnamefont [1]{#1}%
\providecommand \citenamefont [1]{#1}%
\providecommand \href@noop [0]{\@secondoftwo}%
\providecommand \href [0]{\begingroup \@sanitize@url \@href}%
\providecommand \@href[1]{\@@startlink{#1}\@@href}%
\providecommand \@@href[1]{\endgroup#1\@@endlink}%
\providecommand \@sanitize@url [0]{\catcode `\\12\catcode `\$12\catcode
  `\&12\catcode `\#12\catcode `\^12\catcode `\_12\catcode `\%12\relax}%
\providecommand \@@startlink[1]{}%
\providecommand \@@endlink[0]{}%
\providecommand \url  [0]{\begingroup\@sanitize@url \@url }%
\providecommand \@url [1]{\endgroup\@href {#1}{\urlprefix }}%
\providecommand \urlprefix  [0]{URL }%
\providecommand \Eprint [0]{\href }%
\providecommand \doibase [0]{https://doi.org/}%
\providecommand \selectlanguage [0]{\@gobble}%
\providecommand \bibinfo  [0]{\@secondoftwo}%
\providecommand \bibfield  [0]{\@secondoftwo}%
\providecommand \translation [1]{[#1]}%
\providecommand \BibitemOpen [0]{}%
\providecommand \bibitemStop [0]{}%
\providecommand \bibitemNoStop [0]{.\EOS\space}%
\providecommand \EOS [0]{\spacefactor3000\relax}%
\providecommand \BibitemShut  [1]{\csname bibitem#1\endcsname}%
\let\auto@bib@innerbib\@empty
\bibitem [{\citenamefont {Abbott}\ \emph {et~al.}(2019)\citenamefont {Abbott}
  \emph {et~al.}}]{LVK_2019_GWTC1}%
  \BibitemOpen
  \bibfield  {author} {\bibinfo {author} {\bibfnamefont {B.~P.}\ \bibnamefont
  {Abbott}} \emph {et~al.} (\bibinfo {collaboration} {LIGO Scientific
  Collaboration and Virgo Collaboration}),\ }\bibfield  {title} {\bibinfo
  {title} {Gwtc-1: A gravitational-wave transient catalog of compact binary
  mergers observed by ligo and virgo during the first and second observing
  runs},\ }\href {https://doi.org/10.1103/PhysRevX.9.031040} {\bibfield
  {journal} {\bibinfo  {journal} {Phys. Rev. X}\ }\textbf {\bibinfo {volume}
  {9}},\ \bibinfo {pages} {031040} (\bibinfo {year} {2019})},\ \Eprint
  {https://arxiv.org/abs/1811.12907} {arXiv:1811.12907 [astro-ph.HE]}
  \BibitemShut {NoStop}%
\bibitem [{\citenamefont {Abbott}\ \emph {et~al.}(2021)\citenamefont {Abbott}
  \emph {et~al.}}]{LVK_2021_GWTC2}%
  \BibitemOpen
  \bibfield  {author} {\bibinfo {author} {\bibfnamefont {R.}~\bibnamefont
  {Abbott}} \emph {et~al.} (\bibinfo {collaboration} {LIGO Scientific
  Collaboration and Virgo Collaboration}),\ }\bibfield  {title} {\bibinfo
  {title} {Gwtc-2: Compact binary coalescences observed by ligo and virgo
  during the first half of the third observing run},\ }\href
  {https://doi.org/10.1103/PhysRevX.11.021053} {\bibfield  {journal} {\bibinfo
  {journal} {Phys. Rev. X}\ }\textbf {\bibinfo {volume} {11}},\ \bibinfo
  {pages} {021053} (\bibinfo {year} {2021})},\ \Eprint
  {https://arxiv.org/abs/2010.14527} {arXiv:2010.14527 [gr-qc]} \BibitemShut
  {NoStop}%
\bibitem [{\citenamefont {Abbott}\ \emph {et~al.}(2023)\citenamefont {Abbott}
  \emph {et~al.}}]{LVK_2023_GWTC3}%
  \BibitemOpen
  \bibfield  {author} {\bibinfo {author} {\bibfnamefont {R.}~\bibnamefont
  {Abbott}} \emph {et~al.} (\bibinfo {collaboration} {LIGO Scientific
  Collaboration, Virgo Collaboration, and KAGRA Collaboration}),\ }\bibfield
  {title} {\bibinfo {title} {Gwtc-3: Compact binary coalescences observed by
  ligo and virgo during the second part of the third observing run},\ }\href
  {https://doi.org/10.1103/PhysRevX.13.041039} {\bibfield  {journal} {\bibinfo
  {journal} {Phys. Rev. X}\ }\textbf {\bibinfo {volume} {13}},\ \bibinfo
  {pages} {041039} (\bibinfo {year} {2023})},\ \Eprint
  {https://arxiv.org/abs/2111.03606} {arXiv:2111.03606 [gr-qc]} \BibitemShut
  {NoStop}%
\bibitem [{\citenamefont {Abbott}\ \emph {et~al.}(2025)\citenamefont {Abbott}
  \emph {et~al.}}]{LVK_2025_GWTC4}%
  \BibitemOpen
  \bibfield  {author} {\bibinfo {author} {\bibfnamefont {R.}~\bibnamefont
  {Abbott}} \emph {et~al.} (\bibinfo {collaboration} {LIGO Scientific
  Collaboration, Virgo Collaboration, and KAGRA Collaboration}),\ }\bibfield
  {title} {\bibinfo {title} {Gwtc-4.0: Updating the gravitational-wave
  transient catalog with observations from the first part of the fourth
  ligo-virgo-kagra observing run},\ }\href@noop {} {\  (\bibinfo {year}
  {2025})},\ \Eprint {https://arxiv.org/abs/2508.18082} {arXiv:2508.18082
  [gr-qc]} \BibitemShut {NoStop}%
\bibitem [{\citenamefont {Abbott}\ \emph {et~al.}(2020)\citenamefont {Abbott}
  \emph {et~al.}}]{LVK_2020_GW190412}%
  \BibitemOpen
  \bibfield  {author} {\bibinfo {author} {\bibfnamefont {R.}~\bibnamefont
  {Abbott}} \emph {et~al.} (\bibinfo {collaboration} {LIGO Scientific
  Collaboration and Virgo Collaboration}),\ }\bibfield  {title} {\bibinfo
  {title} {Gw190412: Observation of a binary-black-hole coalescence with
  asymmetric masses},\ }\href {https://doi.org/10.1103/PhysRevD.102.043015}
  {\bibfield  {journal} {\bibinfo  {journal} {Phys. Rev. D}\ }\textbf {\bibinfo
  {volume} {102}},\ \bibinfo {pages} {043015} (\bibinfo {year} {2020})},\
  \Eprint {https://arxiv.org/abs/2004.08342} {arXiv:2004.08342 [astro-ph.HE]}
  \BibitemShut {NoStop}%
\bibitem [{\citenamefont {Bustillo}\ \emph {et~al.}(2025)\citenamefont
  {Bustillo}, \citenamefont {Leong},\ and\ \citenamefont
  {Chandra}}]{Bustillo_2022_GW190412}%
  \BibitemOpen
  \bibfield  {author} {\bibinfo {author} {\bibfnamefont {J.~C.}\ \bibnamefont
  {Bustillo}}, \bibinfo {author} {\bibfnamefont {S.~H.~W.}\ \bibnamefont
  {Leong}},\ and\ \bibinfo {author} {\bibfnamefont {K.}~\bibnamefont
  {Chandra}},\ }\bibfield  {title} {\bibinfo {title} {Gw190412: measuring a
  black-hole recoil direction through higher-order gravitational-wave modes},\
  }\href {https://doi.org/10.1038/s41550-025-02632-5} {\bibfield  {journal}
  {\bibinfo  {journal} {Nature Astron.}\ }\textbf {\bibinfo {volume} {9}},\
  \bibinfo {pages} {1530} (\bibinfo {year} {2025})},\ \Eprint
  {https://arxiv.org/abs/2211.03465} {arXiv:2211.03465 [gr-qc]} \BibitemShut
  {NoStop}%
\bibitem [{\citenamefont {Abbott}\ \emph {et~al.}()\citenamefont {Abbott} \emph
  {et~al.}}]{LVK_2020_GW190814}%
  \BibitemOpen
  \bibfield  {author} {\bibinfo {author} {\bibfnamefont {R.}~\bibnamefont
  {Abbott}} \emph {et~al.} (\bibinfo {collaboration} {LIGO Scientific
  Collaboration and Virgo Collaboration}),\ }\bibfield  {title} {\bibinfo
  {title} {Gw190814: Gravitational waves from the coalescence of a 23 solar
  mass black hole with a 2.6 solar mass compact object},\ }\href@noop {}
  {\bibfield  {journal} {\bibinfo  {journal} {Astrophys. J. Lett.}\ }\textbf
  {\bibinfo {volume} {896}},\ \bibinfo {pages} {L44}}\BibitemShut {NoStop}%
\bibitem [{\citenamefont {Abac}\ \emph
  {et~al.}(2025{\natexlab{a}})\citenamefont {Abac} \emph
  {et~al.}}]{LVK_2025_GW241011}%
  \BibitemOpen
  \bibfield  {author} {\bibinfo {author} {\bibfnamefont {A.~G.}\ \bibnamefont
  {Abac}} \emph {et~al.} (\bibinfo {collaboration} {LIGO Scientific
  Collaboration, Virgo Collaboration, and KAGRA Collaboration}),\ }\bibfield
  {title} {\bibinfo {title} {Gw241011 and gw241110: Exploring binary formation
  and fundamental physics with asymmetric, high-spin black hole coalescences},\
  }\href {https://doi.org/10.3847/2041-8213/ae0d54} {\bibfield  {journal}
  {\bibinfo  {journal} {Astrophys. J. Lett.}\ }\textbf {\bibinfo {volume}
  {993}},\ \bibinfo {pages} {L21} (\bibinfo {year} {2025}{\natexlab{a}})},\
  \Eprint {https://arxiv.org/abs/2510.26931} {arXiv:2510.26931 [astro-ph.HE]}
  \BibitemShut {NoStop}%
\bibitem [{\citenamefont {Hannam}\ \emph {et~al.}(2022)\citenamefont {Hannam}
  \emph {et~al.}}]{Hannam_2022_GW200129}%
  \BibitemOpen
  \bibfield  {author} {\bibinfo {author} {\bibfnamefont {M.}~\bibnamefont
  {Hannam}} \emph {et~al.},\ }\bibfield  {title} {\bibinfo {title}
  {General-relativistic precession in a black-hole binary},\ }\href
  {https://doi.org/10.1038/s41586-022-05212-z} {\bibfield  {journal} {\bibinfo
  {journal} {Nature}\ }\textbf {\bibinfo {volume} {610}},\ \bibinfo {pages}
  {652} (\bibinfo {year} {2022})},\ \Eprint {https://arxiv.org/abs/2112.11300}
  {arXiv:2112.11300 [gr-qc]} \BibitemShut {NoStop}%
\bibitem [{\citenamefont {Varma}\ \emph {et~al.}(2022)\citenamefont {Varma},
  \citenamefont {Biscoveanu}, \citenamefont {Islam}, \citenamefont {Shaik},
  \citenamefont {Haster}, \citenamefont {Isi}, \citenamefont {Farr},
  \citenamefont {Field},\ and\ \citenamefont {Vitale}}]{Varma_2022_GW200129}%
  \BibitemOpen
  \bibfield  {author} {\bibinfo {author} {\bibfnamefont {V.}~\bibnamefont
  {Varma}}, \bibinfo {author} {\bibfnamefont {S.}~\bibnamefont {Biscoveanu}},
  \bibinfo {author} {\bibfnamefont {T.}~\bibnamefont {Islam}}, \bibinfo
  {author} {\bibfnamefont {F.~H.}\ \bibnamefont {Shaik}}, \bibinfo {author}
  {\bibfnamefont {C.-J.}\ \bibnamefont {Haster}}, \bibinfo {author}
  {\bibfnamefont {M.}~\bibnamefont {Isi}}, \bibinfo {author} {\bibfnamefont
  {W.~M.}\ \bibnamefont {Farr}}, \bibinfo {author} {\bibfnamefont {S.~E.}\
  \bibnamefont {Field}},\ and\ \bibinfo {author} {\bibfnamefont
  {S.}~\bibnamefont {Vitale}},\ }\bibfield  {title} {\bibinfo {title} {Evidence
  of large recoil velocity from a black hole merger signal},\ }\href
  {https://doi.org/10.1103/PhysRevLett.128.191102} {\bibfield  {journal}
  {\bibinfo  {journal} {Phys. Rev. Lett.}\ }\textbf {\bibinfo {volume} {128}},\
  \bibinfo {pages} {191102} (\bibinfo {year} {2022})},\ \Eprint
  {https://arxiv.org/abs/2201.01302} {arXiv:2201.01302 [gr-qc]} \BibitemShut
  {NoStop}%
\bibitem [{\citenamefont {Kolitsidou}\ \emph {et~al.}(2025)\citenamefont
  {Kolitsidou}, \citenamefont {Thompson},\ and\ \citenamefont
  {Hannam}}]{Kolitsidou_2024_AsymSpinMeasure}%
  \BibitemOpen
  \bibfield  {author} {\bibinfo {author} {\bibfnamefont {P.}~\bibnamefont
  {Kolitsidou}}, \bibinfo {author} {\bibfnamefont {J.~E.}\ \bibnamefont
  {Thompson}},\ and\ \bibinfo {author} {\bibfnamefont {M.}~\bibnamefont
  {Hannam}},\ }\bibfield  {title} {\bibinfo {title} {Impact of anti-symmetric
  contributions to signal multipoles in the measurement of black-hole spins},\
  }\href {https://doi.org/10.1103/PhysRevD.111.024050} {\bibfield  {journal}
  {\bibinfo  {journal} {Phys. Rev. D}\ }\textbf {\bibinfo {volume} {111}},\
  \bibinfo {pages} {024050} (\bibinfo {year} {2025})},\ \Eprint
  {https://arxiv.org/abs/2402.00813} {arXiv:2402.00813 [gr-qc]} \BibitemShut
  {NoStop}%
\bibitem [{\citenamefont {Borchers}\ \emph {et~al.}(2024)\citenamefont
  {Borchers}, \citenamefont {Ohme}, \citenamefont {Mielke},\ and\ \citenamefont
  {Ghosh}}]{Borchers_2024_ObsPrecKick}%
  \BibitemOpen
  \bibfield  {author} {\bibinfo {author} {\bibfnamefont {A.}~\bibnamefont
  {Borchers}}, \bibinfo {author} {\bibfnamefont {F.}~\bibnamefont {Ohme}},
  \bibinfo {author} {\bibfnamefont {J.}~\bibnamefont {Mielke}},\ and\ \bibinfo
  {author} {\bibfnamefont {S.}~\bibnamefont {Ghosh}},\ }\bibfield  {title}
  {\bibinfo {title} {Observability of spin precession in the presence of a
  black-hole remnant kick},\ }\href
  {https://doi.org/10.1103/PhysRevD.110.024037} {\bibfield  {journal} {\bibinfo
   {journal} {Phys. Rev. D}\ }\textbf {\bibinfo {volume} {110}},\ \bibinfo
  {pages} {024037} (\bibinfo {year} {2024})},\ \Eprint
  {https://arxiv.org/abs/2405.03607} {arXiv:2405.03607 [gr-qc]} \BibitemShut
  {NoStop}%
\bibitem [{\citenamefont {Apostolatos}\ \emph {et~al.}(1994)\citenamefont
  {Apostolatos}, \citenamefont {Cutler}, \citenamefont {Sussman},\ and\
  \citenamefont {Thorne}}]{Apostolatos_1994_Precession}%
  \BibitemOpen
  \bibfield  {author} {\bibinfo {author} {\bibfnamefont {T.~A.}\ \bibnamefont
  {Apostolatos}}, \bibinfo {author} {\bibfnamefont {C.}~\bibnamefont {Cutler}},
  \bibinfo {author} {\bibfnamefont {G.~J.}\ \bibnamefont {Sussman}},\ and\
  \bibinfo {author} {\bibfnamefont {K.~S.}\ \bibnamefont {Thorne}},\ }\bibfield
   {title} {\bibinfo {title} {Spin-induced orbital precession and its
  modulation of the gravitational waveforms from merging binaries},\ }\href
  {https://doi.org/10.1103/PhysRevD.49.6274} {\bibfield  {journal} {\bibinfo
  {journal} {Phys. Rev. D}\ }\textbf {\bibinfo {volume} {49}},\ \bibinfo
  {pages} {6274} (\bibinfo {year} {1994})}\BibitemShut {NoStop}%
\bibitem [{\citenamefont {Kidder}(1995)}]{Kidder_1995_Mark52PN}%
  \BibitemOpen
  \bibfield  {author} {\bibinfo {author} {\bibfnamefont {L.~E.}\ \bibnamefont
  {Kidder}},\ }\bibfield  {title} {\bibinfo {title} {Coalescing binary systems
  of compact objects to post-5/2-newtonian order. v. spin effects},\ }\href
  {https://doi.org/10.1103/physrevd.52.821} {\bibfield  {journal} {\bibinfo
  {journal} {Phys. Rev. D}\ }\textbf {\bibinfo {volume} {52}},\ \bibinfo
  {pages} {821} (\bibinfo {year} {1995})},\ \Eprint
  {https://arxiv.org/abs/gr-qc/9506022} {arXiv:gr-qc/9506022} \BibitemShut
  {NoStop}%
\bibitem [{\citenamefont {Pretorius}(2007)}]{Pretorius_2007_BBHCoalescence}%
  \BibitemOpen
  \bibfield  {author} {\bibinfo {author} {\bibfnamefont {F.}~\bibnamefont
  {Pretorius}},\ }\href@noop {} {\bibinfo {title} {Binary black hole
  coalescence}} (\bibinfo {year} {2007}),\ \Eprint
  {https://arxiv.org/abs/0710.1338} {arXiv:0710.1338 [gr-qc]} \BibitemShut
  {NoStop}%
\bibitem [{\citenamefont {Br\"ugmann}\ \emph {et~al.}(2008)\citenamefont
  {Br\"ugmann}, \citenamefont {Gonz\'alez}, \citenamefont {Hannam},
  \citenamefont {Husa},\ and\ \citenamefont
  {Sperhake}}]{Bruegmann_2007_Superkicks}%
  \BibitemOpen
  \bibfield  {author} {\bibinfo {author} {\bibfnamefont {B.}~\bibnamefont
  {Br\"ugmann}}, \bibinfo {author} {\bibfnamefont {J.~A.}\ \bibnamefont
  {Gonz\'alez}}, \bibinfo {author} {\bibfnamefont {M.}~\bibnamefont {Hannam}},
  \bibinfo {author} {\bibfnamefont {S.}~\bibnamefont {Husa}},\ and\ \bibinfo
  {author} {\bibfnamefont {U.}~\bibnamefont {Sperhake}},\ }\bibfield  {title}
  {\bibinfo {title} {Exploring black hole superkicks},\ }\href
  {https://doi.org/10.1103/PhysRevD.77.124047} {\bibfield  {journal} {\bibinfo
  {journal} {Phys. Rev. D}\ }\textbf {\bibinfo {volume} {77}},\ \bibinfo
  {pages} {124047} (\bibinfo {year} {2008})},\ \Eprint
  {https://arxiv.org/abs/0707.0135} {arXiv:0707.0135 [gr-qc]} \BibitemShut
  {NoStop}%
\bibitem [{\citenamefont {Campanelli}\ \emph {et~al.}(2007)\citenamefont
  {Campanelli}, \citenamefont {Lousto}, \citenamefont {Zlochower},\ and\
  \citenamefont {Merritt}}]{Campanelli_2007_superkick}%
  \BibitemOpen
  \bibfield  {author} {\bibinfo {author} {\bibfnamefont {M.}~\bibnamefont
  {Campanelli}}, \bibinfo {author} {\bibfnamefont {C.~O.}\ \bibnamefont
  {Lousto}}, \bibinfo {author} {\bibfnamefont {Y.}~\bibnamefont {Zlochower}},\
  and\ \bibinfo {author} {\bibfnamefont {D.}~\bibnamefont {Merritt}},\
  }\bibfield  {title} {\bibinfo {title} {Maximum gravitational recoil},\ }\href
  {https://doi.org/10.1103/PhysRevLett.98.231102} {\bibfield  {journal}
  {\bibinfo  {journal} {Phys. Rev. Lett.}\ }\textbf {\bibinfo {volume} {98}},\
  \bibinfo {pages} {231102} (\bibinfo {year} {2007})},\ \Eprint
  {https://arxiv.org/abs/gr-qc/0702133} {arXiv:gr-qc/0702133} \BibitemShut
  {NoStop}%
\bibitem [{\citenamefont {Gonz\'alez}\ \emph {et~al.}(2007)\citenamefont
  {Gonz\'alez}, \citenamefont {Hannam}, \citenamefont {Sperhake}, \citenamefont
  {Br\"ugmann},\ and\ \citenamefont {Husa}}]{Gonzalez_2007_superkick}%
  \BibitemOpen
  \bibfield  {author} {\bibinfo {author} {\bibfnamefont {J.~A.}\ \bibnamefont
  {Gonz\'alez}}, \bibinfo {author} {\bibfnamefont {M.}~\bibnamefont {Hannam}},
  \bibinfo {author} {\bibfnamefont {U.}~\bibnamefont {Sperhake}}, \bibinfo
  {author} {\bibfnamefont {B.}~\bibnamefont {Br\"ugmann}},\ and\ \bibinfo
  {author} {\bibfnamefont {S.}~\bibnamefont {Husa}},\ }\bibfield  {title}
  {\bibinfo {title} {Supermassive recoil velocities for binary black-hole
  mergers with antialigned spins},\ }\href
  {https://doi.org/10.1103/PhysRevLett.98.231101} {\bibfield  {journal}
  {\bibinfo  {journal} {Phys. Rev. Lett.}\ }\textbf {\bibinfo {volume} {98}},\
  \bibinfo {pages} {231101} (\bibinfo {year} {2007})},\ \Eprint
  {https://arxiv.org/abs/gr-qc/0702052} {arXiv:gr-qc/0702052} \BibitemShut
  {NoStop}%
\bibitem [{\citenamefont {Ruiz}\ \emph {et~al.}(2007)\citenamefont {Ruiz},
  \citenamefont {Alcubierre}, \citenamefont {N{\'{u}}{\~{n}}ez},\ and\
  \citenamefont {Takahashi}}]{Ruiz_2007_MomentaFluxes}%
  \BibitemOpen
  \bibfield  {author} {\bibinfo {author} {\bibfnamefont {M.}~\bibnamefont
  {Ruiz}}, \bibinfo {author} {\bibfnamefont {M.}~\bibnamefont {Alcubierre}},
  \bibinfo {author} {\bibfnamefont {D.}~\bibnamefont {N{\'{u}}{\~{n}}ez}},\
  and\ \bibinfo {author} {\bibfnamefont {R.}~\bibnamefont {Takahashi}},\
  }\bibfield  {title} {\bibinfo {title} {Multipole expansions for energy and
  momenta carried by gravitational waves},\ }\href
  {https://doi.org/10.1007/s10714-007-0570-8} {\bibfield  {journal} {\bibinfo
  {journal} {Gen. Rel. Grav.}\ }\textbf {\bibinfo {volume} {40}},\ \bibinfo
  {pages} {1705} (\bibinfo {year} {2007})},\ \Eprint
  {https://arxiv.org/abs/0707.4654} {arXiv:0707.4654 [gr-qc]} \BibitemShut
  {NoStop}%
\bibitem [{\citenamefont {Mielke}\ \emph {et~al.}(2025)\citenamefont {Mielke},
  \citenamefont {Ghosh}, \citenamefont {Borchers},\ and\ \citenamefont
  {Ohme}}]{Mielke_2025_kickasym}%
  \BibitemOpen
  \bibfield  {author} {\bibinfo {author} {\bibfnamefont {J.}~\bibnamefont
  {Mielke}}, \bibinfo {author} {\bibfnamefont {S.}~\bibnamefont {Ghosh}},
  \bibinfo {author} {\bibfnamefont {A.}~\bibnamefont {Borchers}},\ and\
  \bibinfo {author} {\bibfnamefont {F.}~\bibnamefont {Ohme}},\ }\bibfield
  {title} {\bibinfo {title} {Revisiting the relationship of black-hole kicks
  and multipole asymmetries},\ }\href
  {https://doi.org/10.1103/PhysRevD.111.064009} {\bibfield  {journal} {\bibinfo
   {journal} {Phys. Rev. D}\ }\textbf {\bibinfo {volume} {111}},\ \bibinfo
  {pages} {064009} (\bibinfo {year} {2025})},\ \Eprint
  {https://arxiv.org/abs/2412.06913} {arXiv:2412.06913 [gr-qc]} \BibitemShut
  {NoStop}%
\bibitem [{\citenamefont {Gerosa}\ \emph {et~al.}(2013)\citenamefont {Gerosa},
  \citenamefont {Kesden}, \citenamefont {Berti}, \citenamefont
  {O'Shaughnessy},\ and\ \citenamefont
  {Sperhake}}]{Gerosa_2013_formationchannel1}%
  \BibitemOpen
  \bibfield  {author} {\bibinfo {author} {\bibfnamefont {D.}~\bibnamefont
  {Gerosa}}, \bibinfo {author} {\bibfnamefont {M.}~\bibnamefont {Kesden}},
  \bibinfo {author} {\bibfnamefont {E.}~\bibnamefont {Berti}}, \bibinfo
  {author} {\bibfnamefont {R.}~\bibnamefont {O'Shaughnessy}},\ and\ \bibinfo
  {author} {\bibfnamefont {U.}~\bibnamefont {Sperhake}},\ }\bibfield  {title}
  {\bibinfo {title} {Resonant-plane locking and spin alignment in stellar-mass
  black-hole binaries: a diagnostic of compact-binary formation},\ }\href
  {https://doi.org/10.1103/PhysRevD.87.104028} {\bibfield  {journal} {\bibinfo
  {journal} {Phys. Rev. D}\ }\textbf {\bibinfo {volume} {87}},\ \bibinfo
  {pages} {104028} (\bibinfo {year} {2013})},\ \Eprint
  {https://arxiv.org/abs/1302.4442} {arXiv:1302.4442 [gr-qc]} \BibitemShut
  {NoStop}%
\bibitem [{\citenamefont {Vitale}\ \emph {et~al.}(2017)\citenamefont {Vitale},
  \citenamefont {Lynch}, \citenamefont {Sturani},\ and\ \citenamefont
  {Graff}}]{Vitale_2017_formationchannel2}%
  \BibitemOpen
  \bibfield  {author} {\bibinfo {author} {\bibfnamefont {S.}~\bibnamefont
  {Vitale}}, \bibinfo {author} {\bibfnamefont {R.}~\bibnamefont {Lynch}},
  \bibinfo {author} {\bibfnamefont {R.}~\bibnamefont {Sturani}},\ and\ \bibinfo
  {author} {\bibfnamefont {P.}~\bibnamefont {Graff}},\ }\bibfield  {title}
  {\bibinfo {title} {Use of gravitational waves to probe the formation channels
  of compact binaries},\ }\href {https://doi.org/10.1088/1361-6382/aa552e}
  {\bibfield  {journal} {\bibinfo  {journal} {Class. Quant. Grav.}\ }\textbf
  {\bibinfo {volume} {34}},\ \bibinfo {pages} {03LT01} (\bibinfo {year}
  {2017})},\ \Eprint {https://arxiv.org/abs/1503.04307} {arXiv:1503.04307
  [gr-qc]} \BibitemShut {NoStop}%
\bibitem [{\citenamefont {Farr}\ \emph {et~al.}(2018)\citenamefont {Farr},
  \citenamefont {Holz},\ and\ \citenamefont
  {Farr}}]{Farr_2018_formationchannel3}%
  \BibitemOpen
  \bibfield  {author} {\bibinfo {author} {\bibfnamefont {B.}~\bibnamefont
  {Farr}}, \bibinfo {author} {\bibfnamefont {D.~E.}\ \bibnamefont {Holz}},\
  and\ \bibinfo {author} {\bibfnamefont {W.~M.}\ \bibnamefont {Farr}},\
  }\bibfield  {title} {\bibinfo {title} {Using spin to understand the formation
  of ligo's black holes},\ }\href {https://doi.org/10.3847/2041-8213/aaaa64}
  {\bibfield  {journal} {\bibinfo  {journal} {Astrphys. J.}\ }\textbf {\bibinfo
  {volume} {854}},\ \bibinfo {pages} {L9} (\bibinfo {year} {2018})},\ \Eprint
  {https://arxiv.org/abs/1709.07896} {arXiv:1709.07896 [astro-ph.HE]}
  \BibitemShut {NoStop}%
\bibitem [{\citenamefont {Mapelli}(2021)}]{Mapelli_2021_formationchannel4}%
  \BibitemOpen
  \bibfield  {author} {\bibinfo {author} {\bibfnamefont {M.}~\bibnamefont
  {Mapelli}},\ }\bibfield  {title} {\bibinfo {title} {Formation channels of
  single and binary stellar-mass black holes},\ }\href@noop {} {\  (\bibinfo
  {year} {2021})},\ \Eprint {https://arxiv.org/abs/2106.00699}
  {arXiv:2106.00699 [astro-ph.HE]} \BibitemShut {NoStop}%
\bibitem [{\citenamefont {Gerosa}\ and\ \citenamefont
  {Fishbach}(2021{\natexlab{a}})}]{Gerosa_2021_popreview}%
  \BibitemOpen
  \bibfield  {author} {\bibinfo {author} {\bibfnamefont {D.}~\bibnamefont
  {Gerosa}}\ and\ \bibinfo {author} {\bibfnamefont {M.}~\bibnamefont
  {Fishbach}},\ }\bibfield  {title} {\bibinfo {title} {Hierarchical mergers of
  stellar-mass black holes and their gravitational-wave signatures},\ }\href
  {https://doi.org/10.1038/s41550-021-01398-w} {\bibfield  {journal} {\bibinfo
  {journal} {Nature Astron.}\ }\textbf {\bibinfo {volume} {5}},\ \bibinfo
  {pages} {749} (\bibinfo {year} {2021}{\natexlab{a}})},\ \Eprint
  {https://arxiv.org/abs/2105.03439} {arXiv:2105.03439 [astro-ph.HE]}
  \BibitemShut {NoStop}%
\bibitem [{\citenamefont {Mahapatra}\ \emph {et~al.}(2021)\citenamefont
  {Mahapatra}, \citenamefont {Gupta}, \citenamefont {Favata}, \citenamefont
  {Arun},\ and\ \citenamefont
  {Sathyaprakash}}]{Mahapatra_2021_HierachicalMergers}%
  \BibitemOpen
  \bibfield  {author} {\bibinfo {author} {\bibfnamefont {P.}~\bibnamefont
  {Mahapatra}}, \bibinfo {author} {\bibfnamefont {A.}~\bibnamefont {Gupta}},
  \bibinfo {author} {\bibfnamefont {M.}~\bibnamefont {Favata}}, \bibinfo
  {author} {\bibfnamefont {K.~G.}\ \bibnamefont {Arun}},\ and\ \bibinfo
  {author} {\bibfnamefont {B.~S.}\ \bibnamefont {Sathyaprakash}},\ }\bibfield
  {title} {\bibinfo {title} {Remnant black hole kicks and implications for
  hierarchical mergers},\ }\href {https://doi.org/10.3847/2041-8213/ac20db}
  {\bibfield  {journal} {\bibinfo  {journal} {Astrophys. J. Lett.}\ }\textbf
  {\bibinfo {volume} {918}},\ \bibinfo {pages} {L31} (\bibinfo {year}
  {2021})},\ \Eprint {https://arxiv.org/abs/2106.07179} {arXiv:2106.07179
  [astro-ph.HE]} \BibitemShut {NoStop}%
\bibitem [{\citenamefont {Araújo-Álvarez}\ \emph {et~al.}(2024)\citenamefont
  {Araújo-Álvarez}, \citenamefont {Wong}, \citenamefont {Liu},\ and\
  \citenamefont {Calderón~Bustillo}}]{AraujoAlvarez_2024_HierachicalGW190521}%
  \BibitemOpen
  \bibfield  {author} {\bibinfo {author} {\bibfnamefont {C.}~\bibnamefont
  {Araújo-Álvarez}}, \bibinfo {author} {\bibfnamefont {H.~W.~Y.}\
  \bibnamefont {Wong}}, \bibinfo {author} {\bibfnamefont {A.}~\bibnamefont
  {Liu}},\ and\ \bibinfo {author} {\bibfnamefont {J.}~\bibnamefont
  {Calderón~Bustillo}},\ }\bibfield  {title} {\bibinfo {title} {Kicking time
  back in black hole mergers: Ancestral masses, spins, birth recoils, and
  hierarchical-formation viability of gw190521},\ }\href
  {https://doi.org/10.3847/1538-4357/ad90a9} {\bibfield  {journal} {\bibinfo
  {journal} {Astrophys. J.}\ }\textbf {\bibinfo {volume} {977}},\ \bibinfo
  {pages} {220} (\bibinfo {year} {2024})},\ \Eprint
  {https://arxiv.org/abs/2404.00720} {arXiv:2404.00720 [astro-ph.HE]}
  \BibitemShut {NoStop}%
\bibitem [{\citenamefont {Mahapatra}\ \emph {et~al.}(2024)\citenamefont
  {Mahapatra}, \citenamefont {Chattopadhyay}, \citenamefont {Gupta},
  \citenamefont {Antonini}, \citenamefont {Favata}, \citenamefont
  {Sathyaprakash},\ and\ \citenamefont {Arun}}]{Mahapatra_2024_Genealogy}%
  \BibitemOpen
  \bibfield  {author} {\bibinfo {author} {\bibfnamefont {P.}~\bibnamefont
  {Mahapatra}}, \bibinfo {author} {\bibfnamefont {D.}~\bibnamefont
  {Chattopadhyay}}, \bibinfo {author} {\bibfnamefont {A.}~\bibnamefont
  {Gupta}}, \bibinfo {author} {\bibfnamefont {F.}~\bibnamefont {Antonini}},
  \bibinfo {author} {\bibfnamefont {M.}~\bibnamefont {Favata}}, \bibinfo
  {author} {\bibfnamefont {B.~S.}\ \bibnamefont {Sathyaprakash}},\ and\
  \bibinfo {author} {\bibfnamefont {K.~G.}\ \bibnamefont {Arun}},\ }\bibfield
  {title} {\bibinfo {title} {Reconstructing the genealogy of ligo-virgo black
  holes},\ }\href {https://doi.org/10.3847/1538-4357/ad781b} {\bibfield
  {journal} {\bibinfo  {journal} {Astrophys. J.}\ }\textbf {\bibinfo {volume}
  {975}},\ \bibinfo {pages} {117} (\bibinfo {year} {2024})},\ \Eprint
  {https://arxiv.org/abs/2406.06390} {arXiv:2406.06390 [astro-ph.HE]}
  \BibitemShut {NoStop}%
\bibitem [{\citenamefont {Borchers}\ \emph {et~al.}(2025)\citenamefont
  {Borchers}, \citenamefont {Ye},\ and\ \citenamefont
  {Fishbach}}]{Borchers_2025_HierMerg}%
  \BibitemOpen
  \bibfield  {author} {\bibinfo {author} {\bibfnamefont {A.}~\bibnamefont
  {Borchers}}, \bibinfo {author} {\bibfnamefont {C.~S.}\ \bibnamefont {Ye}},\
  and\ \bibinfo {author} {\bibfnamefont {M.}~\bibnamefont {Fishbach}},\
  }\bibfield  {title} {\bibinfo {title} {Gravitational-wave kicks impact spins
  of black holes from hierarchical mergers},\ }\href@noop {} {\bibfield
  {journal} {\bibinfo  {journal} {Astrophys. J.}\ }\textbf {\bibinfo {volume}
  {987}},\ \bibinfo {pages} {146} (\bibinfo {year} {2025})},\ \Eprint
  {https://arxiv.org/abs/2503.21278} {arXiv:2503.21278 [astro-ph.HE]}
  \BibitemShut {NoStop}%
\bibitem [{\citenamefont {Leong}\ and\ \citenamefont
  {Bustillo}(2025)}]{Leong_2025_KickMultiMessenger}%
  \BibitemOpen
  \bibfield  {author} {\bibinfo {author} {\bibfnamefont {S.~H.~W.}\
  \bibnamefont {Leong}}\ and\ \bibinfo {author} {\bibfnamefont {J.~C.}\
  \bibnamefont {Bustillo}},\ }\bibfield  {title} {\bibinfo {title} {Kick and
  spin: new probes for multi-messenger black-hole mergers in agns},\
  }\href@noop {} {\  (\bibinfo {year} {2025})},\ \Eprint
  {https://arxiv.org/abs/2512.08382} {arXiv:2512.08382 [astro-ph.HE]}
  \BibitemShut {NoStop}%
\bibitem [{\citenamefont {Graham}\ \emph {et~al.}(2023)\citenamefont {Graham}
  \emph {et~al.}}]{Graham_2022_BBHcounterparts}%
  \BibitemOpen
  \bibfield  {author} {\bibinfo {author} {\bibfnamefont {M.~J.}\ \bibnamefont
  {Graham}} \emph {et~al.},\ }\bibfield  {title} {\bibinfo {title} {A light in
  the dark: Searching for electromagnetic counterparts to black hole-black hole
  mergers in ligo/virgo o3 with the zwicky transient facility},\ }\href
  {https://doi.org/10.3847/1538-4357/aca480} {\bibfield  {journal} {\bibinfo
  {journal} {Astrophys. J.}\ }\textbf {\bibinfo {volume} {942}},\ \bibinfo
  {pages} {99} (\bibinfo {year} {2023})},\ \Eprint
  {https://arxiv.org/abs/2209.13004} {arXiv:2209.13004 [astro-ph.HE]}
  \BibitemShut {NoStop}%
\bibitem [{\citenamefont {Shaikh}\ \emph {et~al.}(2025)\citenamefont {Shaikh},
  \citenamefont {Varma}, \citenamefont {Ramos-Buades}, \citenamefont
  {Pfeiffer}, \citenamefont {Boyle}, \citenamefont {Kidder},\ and\
  \citenamefont {Scheel}}]{Shaikh_2025_EccDef}%
  \BibitemOpen
  \bibfield  {author} {\bibinfo {author} {\bibfnamefont {M.~A.}\ \bibnamefont
  {Shaikh}}, \bibinfo {author} {\bibfnamefont {V.}~\bibnamefont {Varma}},
  \bibinfo {author} {\bibfnamefont {A.}~\bibnamefont {Ramos-Buades}}, \bibinfo
  {author} {\bibfnamefont {H.~P.}\ \bibnamefont {Pfeiffer}}, \bibinfo {author}
  {\bibfnamefont {M.}~\bibnamefont {Boyle}}, \bibinfo {author} {\bibfnamefont
  {L.~E.}\ \bibnamefont {Kidder}},\ and\ \bibinfo {author} {\bibfnamefont
  {M.~A.}\ \bibnamefont {Scheel}},\ }\bibfield  {title} {\bibinfo {title}
  {Defining eccentricity for spin-precessing binaries},\ }\href@noop {} {\
  (\bibinfo {year} {2025})},\ \Eprint {https://arxiv.org/abs/2507.08345}
  {arXiv:2507.08345 [gr-qc]} \BibitemShut {NoStop}%
\bibitem [{\citenamefont {Thompson}\ \emph {et~al.}(2024)\citenamefont
  {Thompson}, \citenamefont {Hamilton}, \citenamefont {London}, \citenamefont
  {Ghosh}, \citenamefont {Kolitsidou}, \citenamefont {Hoy},\ and\ \citenamefont
  {Hannam}}]{Thompson_2024_IMRPhenomXO4a}%
  \BibitemOpen
  \bibfield  {author} {\bibinfo {author} {\bibfnamefont {J.~E.}\ \bibnamefont
  {Thompson}}, \bibinfo {author} {\bibfnamefont {E.}~\bibnamefont {Hamilton}},
  \bibinfo {author} {\bibfnamefont {L.}~\bibnamefont {London}}, \bibinfo
  {author} {\bibfnamefont {S.}~\bibnamefont {Ghosh}}, \bibinfo {author}
  {\bibfnamefont {P.}~\bibnamefont {Kolitsidou}}, \bibinfo {author}
  {\bibfnamefont {C.}~\bibnamefont {Hoy}},\ and\ \bibinfo {author}
  {\bibfnamefont {M.}~\bibnamefont {Hannam}},\ }\bibfield  {title} {\bibinfo
  {title} {Phenomenological gravitational-wave model for precessing black-hole
  binaries with higher multipoles and asymmetries},\ }\href
  {https://doi.org/10.1103/physrevd.109.063012} {\bibfield  {journal} {\bibinfo
   {journal} {Phys. Rev. D}\ }\textbf {\bibinfo {volume} {109}},\ \bibinfo
  {pages} {063012} (\bibinfo {year} {2024})},\ \Eprint
  {https://arxiv.org/abs/2312.10025} {arXiv:2312.10025 [gr-qc]} \BibitemShut
  {NoStop}%
\bibitem [{\citenamefont {Ghosh}\ \emph {et~al.}(2024)\citenamefont {Ghosh},
  \citenamefont {Kolitsidou},\ and\ \citenamefont
  {Hannam}}]{Ghosh_2024_ModeAsymmetryXO4a}%
  \BibitemOpen
  \bibfield  {author} {\bibinfo {author} {\bibfnamefont {S.}~\bibnamefont
  {Ghosh}}, \bibinfo {author} {\bibfnamefont {P.}~\bibnamefont {Kolitsidou}},\
  and\ \bibinfo {author} {\bibfnamefont {M.}~\bibnamefont {Hannam}},\
  }\bibfield  {title} {\bibinfo {title} {First frequency-domain
  phenomenological model of the multipole asymmetry in gravitational-wave
  signals from binary-black-hole coalescence},\ }\href
  {https://doi.org/10.1103/PhysRevD.109.024061} {\bibfield  {journal} {\bibinfo
   {journal} {Phys. Rev. D}\ }\textbf {\bibinfo {volume} {109}},\ \bibinfo
  {pages} {024061} (\bibinfo {year} {2024})},\ \Eprint
  {https://arxiv.org/abs/2310.16980} {arXiv:2310.16980 [gr-qc]} \BibitemShut
  {NoStop}%
\bibitem [{\citenamefont {Hamilton}\ \emph {et~al.}(2025)\citenamefont
  {Hamilton} \emph {et~al.}}]{Hamilton_2025_XPNR}%
  \BibitemOpen
  \bibfield  {author} {\bibinfo {author} {\bibfnamefont {E.}~\bibnamefont
  {Hamilton}} \emph {et~al.},\ }\bibfield  {title} {\bibinfo {title}
  {Phenomxpnr: An improved gravitational wave model linking precessing
  inspirals and nr-calibrated merger-ringdown},\ }\href@noop {} {\  (\bibinfo
  {year} {2025})},\ \Eprint {https://arxiv.org/abs/2507.02604}
  {arXiv:2507.02604 [gr-qc]} \BibitemShut {NoStop}%
\bibitem [{\citenamefont {Estellés}\ \emph {et~al.}(2025)\citenamefont
  {Estellés}, \citenamefont {Buonanno}, \citenamefont {Enficiaud},
  \citenamefont {Foo},\ and\ \citenamefont {Pompili}}]{Estelles_2025_EOBasym}%
  \BibitemOpen
  \bibfield  {author} {\bibinfo {author} {\bibfnamefont {H.}~\bibnamefont
  {Estellés}}, \bibinfo {author} {\bibfnamefont {A.}~\bibnamefont {Buonanno}},
  \bibinfo {author} {\bibfnamefont {R.}~\bibnamefont {Enficiaud}}, \bibinfo
  {author} {\bibfnamefont {C.}~\bibnamefont {Foo}},\ and\ \bibinfo {author}
  {\bibfnamefont {L.}~\bibnamefont {Pompili}},\ }\bibfield  {title} {\bibinfo
  {title} {Adding equatorial-asymmetric effects for spin-precessing binaries
  into the seobnrv5phm waveform model},\ }\href@noop {} {\  (\bibinfo {year}
  {2025})},\ \Eprint {https://arxiv.org/abs/2506.19911} {arXiv:2506.19911
  [gr-qc]} \BibitemShut {NoStop}%
\bibitem [{\citenamefont {Varma}\ \emph {et~al.}(2019)\citenamefont {Varma},
  \citenamefont {Field}, \citenamefont {Scheel}, \citenamefont {Blackman},
  \citenamefont {Gerosa}, \citenamefont {Stein}, \citenamefont {Kidder},\ and\
  \citenamefont {Pfeiffer}}]{Varma_2019_NRSur7dq4}%
  \BibitemOpen
  \bibfield  {author} {\bibinfo {author} {\bibfnamefont {V.}~\bibnamefont
  {Varma}}, \bibinfo {author} {\bibfnamefont {S.~E.}\ \bibnamefont {Field}},
  \bibinfo {author} {\bibfnamefont {M.~A.}\ \bibnamefont {Scheel}}, \bibinfo
  {author} {\bibfnamefont {J.}~\bibnamefont {Blackman}}, \bibinfo {author}
  {\bibfnamefont {D.}~\bibnamefont {Gerosa}}, \bibinfo {author} {\bibfnamefont
  {L.~C.}\ \bibnamefont {Stein}}, \bibinfo {author} {\bibfnamefont {L.~E.}\
  \bibnamefont {Kidder}},\ and\ \bibinfo {author} {\bibfnamefont {H.~P.}\
  \bibnamefont {Pfeiffer}},\ }\bibfield  {title} {\bibinfo {title} {Surrogate
  models for precessing binary black hole simulations with unequal masses},\
  }\href {https://doi.org/10.1103/physrevresearch.1.033015} {\bibfield
  {journal} {\bibinfo  {journal} {Phys. Rev. Res.}\ }\textbf {\bibinfo {volume}
  {1}},\ \bibinfo {pages} {033015} (\bibinfo {year} {2019})},\ \Eprint
  {https://arxiv.org/abs/1905.09300} {arXiv:1905.09300 [gr-qc]} \BibitemShut
  {NoStop}%
\bibitem [{\citenamefont {Schmidt}\ \emph {et~al.}(2011)\citenamefont
  {Schmidt}, \citenamefont {Hannam}, \citenamefont {Husa},\ and\ \citenamefont
  {Ajith}}]{Schmidt_2011_QAframe}%
  \BibitemOpen
  \bibfield  {author} {\bibinfo {author} {\bibfnamefont {P.}~\bibnamefont
  {Schmidt}}, \bibinfo {author} {\bibfnamefont {M.}~\bibnamefont {Hannam}},
  \bibinfo {author} {\bibfnamefont {S.}~\bibnamefont {Husa}},\ and\ \bibinfo
  {author} {\bibfnamefont {P.}~\bibnamefont {Ajith}},\ }\bibfield  {title}
  {\bibinfo {title} {Tracking the precession of compact binaries from their
  gravitational-wave signal},\ }\href
  {https://doi.org/10.1103/PhysRevD.84.024046} {\bibfield  {journal} {\bibinfo
  {journal} {Phys. Rev. D}\ }\textbf {\bibinfo {volume} {84}},\ \bibinfo
  {pages} {024046} (\bibinfo {year} {2011})},\ \Eprint
  {https://arxiv.org/abs/1012.2879} {arXiv:1012.2879 [gr-qc]} \BibitemShut
  {NoStop}%
\bibitem [{\citenamefont {Boyle}\ \emph {et~al.}(2011)\citenamefont {Boyle},
  \citenamefont {Owen},\ and\ \citenamefont {Pfeiffer}}]{Boyle_2011_CoprFrame}%
  \BibitemOpen
  \bibfield  {author} {\bibinfo {author} {\bibfnamefont {M.}~\bibnamefont
  {Boyle}}, \bibinfo {author} {\bibfnamefont {R.}~\bibnamefont {Owen}},\ and\
  \bibinfo {author} {\bibfnamefont {H.~P.}\ \bibnamefont {Pfeiffer}},\
  }\bibfield  {title} {\bibinfo {title} {Geometric approach to the precession
  of compact binaries},\ }\href {https://doi.org/10.1103/PhysRevD.84.124011}
  {\bibfield  {journal} {\bibinfo  {journal} {Phys. Rev. D}\ }\textbf {\bibinfo
  {volume} {84}},\ \bibinfo {pages} {124011} (\bibinfo {year} {2011})},\
  \Eprint {https://arxiv.org/abs/1110.2965} {arXiv:1110.2965 [gr-qc]}
  \BibitemShut {NoStop}%
\bibitem [{\citenamefont {Boyle}\ \emph {et~al.}(2014)\citenamefont {Boyle},
  \citenamefont {Kidder}, \citenamefont {Ossokine},\ and\ \citenamefont
  {Pfeiffer}}]{Boyle_2014_PrecessingBHOperators}%
  \BibitemOpen
  \bibfield  {author} {\bibinfo {author} {\bibfnamefont {M.}~\bibnamefont
  {Boyle}}, \bibinfo {author} {\bibfnamefont {L.~E.}\ \bibnamefont {Kidder}},
  \bibinfo {author} {\bibfnamefont {S.}~\bibnamefont {Ossokine}},\ and\
  \bibinfo {author} {\bibfnamefont {H.~P.}\ \bibnamefont {Pfeiffer}},\
  }\bibfield  {title} {\bibinfo {title} {Gravitational-wave modes from
  precessing black-hole binaries},\ }\href@noop {} {\  (\bibinfo {year}
  {2014})},\ \Eprint {https://arxiv.org/abs/1409.4431} {arXiv:1409.4431
  [gr-qc]} \BibitemShut {NoStop}%
\bibitem [{\citenamefont {W.}(1985)}]{Leaver_1985_QNManalysis}%
  \BibitemOpen
  \bibfield  {author} {\bibinfo {author} {\bibfnamefont {L.~E.}\ \bibnamefont
  {W.}},\ }\bibfield  {title} {\bibinfo {title} {An analytic representation for
  the quasi-normal modes of kerr black holes},\ }\href
  {https://doi.org/10.1098/rspa.1985.0119} {\bibfield  {journal} {\bibinfo
  {journal} {Proc. Roy. Soc. Lond. A}\ }\textbf {\bibinfo {volume} {402}},\
  \bibinfo {pages} {285} (\bibinfo {year} {1985})}\BibitemShut {NoStop}%
\bibitem [{\citenamefont {Moreschi}(1988)}]{Moreschi_1988_supercentre}%
  \BibitemOpen
  \bibfield  {author} {\bibinfo {author} {\bibfnamefont {O.~M.}\ \bibnamefont
  {Moreschi}},\ }\bibfield  {title} {\bibinfo {title} {Supercentre of mass
  system at future null infinity},\ }\href
  {https://doi.org/10.1088/0264-9381/5/3/004} {\bibfield  {journal} {\bibinfo
  {journal} {Class. Quant. Grav.}\ }\textbf {\bibinfo {volume} {5}},\ \bibinfo
  {pages} {423} (\bibinfo {year} {1988})}\BibitemShut {NoStop}%
\bibitem [{\citenamefont {Moreschi}\ and\ \citenamefont
  {Dain}(1998)}]{Moreschi_1998_restframe}%
  \BibitemOpen
  \bibfield  {author} {\bibinfo {author} {\bibfnamefont {O.~M.}\ \bibnamefont
  {Moreschi}}\ and\ \bibinfo {author} {\bibfnamefont {S.}~\bibnamefont
  {Dain}},\ }\bibfield  {title} {\bibinfo {title} {Rest frame system for
  asymptotically flat space-times},\ }\href {https://doi.org/10.1063/1.532646}
  {\bibfield  {journal} {\bibinfo  {journal} {J. Math. Phys.}\ }\textbf
  {\bibinfo {volume} {39}},\ \bibinfo {pages} {6631} (\bibinfo {year}
  {1998})},\ \Eprint {https://arxiv.org/abs/gr-qc/0203075}
  {arXiv:gr-qc/0203075} \BibitemShut {NoStop}%
\bibitem [{\citenamefont {Maga\~na Zertuche}\ \emph {et~al.}(2022)\citenamefont
  {Maga\~na Zertuche} \emph {et~al.}}]{Zertuche_2022_BMSFrameQNM}%
  \BibitemOpen
  \bibfield  {author} {\bibinfo {author} {\bibfnamefont {L.}~\bibnamefont
  {Maga\~na Zertuche}} \emph {et~al.},\ }\bibfield  {title} {\bibinfo {title}
  {High precision ringdown modeling: Multimode fits and bms frames},\ }\href
  {https://doi.org/10.1103/PhysRevD.105.104015} {\bibfield  {journal} {\bibinfo
   {journal} {Phys. Rev. D}\ }\textbf {\bibinfo {volume} {105}},\ \bibinfo
  {pages} {104015} (\bibinfo {year} {2022})},\ \Eprint
  {https://arxiv.org/abs/2110.15922} {arXiv:2110.15922 [gr-qc]} \BibitemShut
  {NoStop}%
\bibitem [{\citenamefont {{Goldberg}}\ \emph {et~al.}(1967)\citenamefont
  {{Goldberg}}, \citenamefont {{Macfarlane}}, \citenamefont {{Newman}},
  \citenamefont {{Rohrlich}},\ and\ \citenamefont
  {{Sudarshan}}}]{Goldberg_1967_delSWSH}%
  \BibitemOpen
  \bibfield  {author} {\bibinfo {author} {\bibfnamefont {J.~N.}\ \bibnamefont
  {{Goldberg}}}, \bibinfo {author} {\bibfnamefont {A.~J.}\ \bibnamefont
  {{Macfarlane}}}, \bibinfo {author} {\bibfnamefont {E.}~\bibnamefont
  {{Newman}}}, \bibinfo {author} {\bibfnamefont {F.}~\bibnamefont
  {{Rohrlich}}},\ and\ \bibinfo {author} {\bibfnamefont {E.}~\bibnamefont
  {{Sudarshan}}},\ }\bibfield  {title} {\bibinfo {title} {Spin-s spherical
  harmonics and edth},\ }\href {https://doi.org/10.1063/1.1705135} {\bibfield
  {journal} {\bibinfo  {journal} {J. Math. Phys.}\ }\textbf {\bibinfo {volume}
  {8}},\ \bibinfo {pages} {2155} (\bibinfo {year} {1967})}\BibitemShut
  {NoStop}%
\bibitem [{\citenamefont {Thorne}(1980)}]{Thorne_1980_Multipole}%
  \BibitemOpen
  \bibfield  {author} {\bibinfo {author} {\bibfnamefont {K.~S.}\ \bibnamefont
  {Thorne}},\ }\bibfield  {title} {\bibinfo {title} {Multipole expansions of
  gravitational radiation},\ }\href {https://doi.org/10.1103/RevModPhys.52.299}
  {\bibfield  {journal} {\bibinfo  {journal} {Rev. Mod. Phys.}\ }\textbf
  {\bibinfo {volume} {52}},\ \bibinfo {pages} {299} (\bibinfo {year}
  {1980})}\BibitemShut {NoStop}%
\bibitem [{\citenamefont {Boyle}\ \emph {et~al.}(2020)\citenamefont {Boyle},
  \citenamefont {Iozzo},\ and\ \citenamefont {Stein}}]{Boyle_2020_scri}%
  \BibitemOpen
  \bibfield  {author} {\bibinfo {author} {\bibfnamefont {M.}~\bibnamefont
  {Boyle}}, \bibinfo {author} {\bibfnamefont {D.}~\bibnamefont {Iozzo}},\ and\
  \bibinfo {author} {\bibfnamefont {L.~C.}\ \bibnamefont {Stein}},\ }\href
  {https://doi.org/10.5281/zenodo.4041972} {\bibinfo {title} {moble/scri:
  v1.2}} (\bibinfo {year} {2020})\BibitemShut {NoStop}%
\bibitem [{\citenamefont {Misner}\ \emph {et~al.}(1973)\citenamefont {Misner},
  \citenamefont {Thorne},\ and\ \citenamefont
  {Wheeler}}]{Misner_1973_MTWGravitation}%
  \BibitemOpen
  \bibfield  {author} {\bibinfo {author} {\bibfnamefont {C.~W.}\ \bibnamefont
  {Misner}}, \bibinfo {author} {\bibfnamefont {K.~S.}\ \bibnamefont {Thorne}},\
  and\ \bibinfo {author} {\bibfnamefont {J.~A.}\ \bibnamefont {Wheeler}},\
  }\href@noop {} {\emph {\bibinfo {title} {Gravitation}}},\ \bibinfo {edition}
  {1st}\ ed.\ (\bibinfo  {publisher} {W. H. Freeman},\ \bibinfo {year}
  {1973})\BibitemShut {NoStop}%
\bibitem [{\citenamefont {Racine}\ \emph {et~al.}(2009)\citenamefont {Racine},
  \citenamefont {Buonanno},\ and\ \citenamefont
  {Kidder}}]{Racine_2009_2PNKick}%
  \BibitemOpen
  \bibfield  {author} {\bibinfo {author} {\bibfnamefont {E.}~\bibnamefont
  {Racine}}, \bibinfo {author} {\bibfnamefont {A.}~\bibnamefont {Buonanno}},\
  and\ \bibinfo {author} {\bibfnamefont {L.}~\bibnamefont {Kidder}},\
  }\bibfield  {title} {\bibinfo {title} {Recoil velocity at second
  post-newtonian order for spinning black hole binaries},\ }\href
  {https://doi.org/10.1103/physrevd.80.044010} {\bibfield  {journal} {\bibinfo
  {journal} {Phys. Rev. D}\ }\textbf {\bibinfo {volume} {80}},\ \bibinfo
  {pages} {044010} (\bibinfo {year} {2009})},\ \Eprint
  {https://arxiv.org/abs/0812.4413} {arXiv:0812.4413 [gr-qc]} \BibitemShut
  {NoStop}%
\bibitem [{\citenamefont {Borchers}\ and\ \citenamefont
  {Ohme}(2023)}]{Borchers_2023_kickbasedGWtuning}%
  \BibitemOpen
  \bibfield  {author} {\bibinfo {author} {\bibfnamefont {A.}~\bibnamefont
  {Borchers}}\ and\ \bibinfo {author} {\bibfnamefont {F.}~\bibnamefont
  {Ohme}},\ }\bibfield  {title} {\bibinfo {title} {Inconsistent black hole kick
  estimates from gravitational-wave models},\ }\href
  {https://doi.org/10.1088/1361-6382/acc5da} {\bibfield  {journal} {\bibinfo
  {journal} {Class. Quantum Gravity}\ }\textbf {\bibinfo {volume} {40}},\
  \bibinfo {pages} {095008} (\bibinfo {year} {2023})},\ \Eprint
  {https://arxiv.org/abs/2207.13531} {arXiv:2207.13531 [gr-qc]} \BibitemShut
  {NoStop}%
\bibitem [{\citenamefont {Merritt}\ \emph {et~al.}(2004)\citenamefont
  {Merritt}, \citenamefont {Milosavljevic}, \citenamefont {Favata},
  \citenamefont {Hughes},\ and\ \citenamefont {Holz}}]{Merritt_2004_vesc}%
  \BibitemOpen
  \bibfield  {author} {\bibinfo {author} {\bibfnamefont {D.}~\bibnamefont
  {Merritt}}, \bibinfo {author} {\bibfnamefont {M.}~\bibnamefont
  {Milosavljevic}}, \bibinfo {author} {\bibfnamefont {M.}~\bibnamefont
  {Favata}}, \bibinfo {author} {\bibfnamefont {S.~A.}\ \bibnamefont {Hughes}},\
  and\ \bibinfo {author} {\bibfnamefont {D.~E.}\ \bibnamefont {Holz}},\
  }\bibfield  {title} {\bibinfo {title} {Consequences of gravitational
  radiation recoil},\ }\href {https://doi.org/10.1086/421551} {\bibfield
  {journal} {\bibinfo  {journal} {Astrophys. J. Lett.}\ }\textbf {\bibinfo
  {volume} {607}},\ \bibinfo {pages} {L9} (\bibinfo {year} {2004})},\ \Eprint
  {https://arxiv.org/abs/astro-ph/0402057} {arXiv:astro-ph/0402057}
  \BibitemShut {NoStop}%
\bibitem [{\citenamefont {Gerosa}\ and\ \citenamefont
  {Fishbach}(2021{\natexlab{b}})}]{Gerosa_2021_vesc}%
  \BibitemOpen
  \bibfield  {author} {\bibinfo {author} {\bibfnamefont {D.}~\bibnamefont
  {Gerosa}}\ and\ \bibinfo {author} {\bibfnamefont {M.}~\bibnamefont
  {Fishbach}},\ }\bibfield  {title} {\bibinfo {title} {Hierarchical mergers of
  stellar-mass black holes and their gravitational-wave signatures},\ }\href
  {https://doi.org/10.1038/s41550-021-01398-w} {\bibfield  {journal} {\bibinfo
  {journal} {Nature Astron.}\ }\textbf {\bibinfo {volume} {5}},\ \bibinfo
  {pages} {749} (\bibinfo {year} {2021}{\natexlab{b}})},\ \Eprint
  {https://arxiv.org/abs/2105.03439} {arXiv:2105.03439 [astro-ph.HE]}
  \BibitemShut {NoStop}%
\bibitem [{\citenamefont {Ma}\ \emph {et~al.}(2021)\citenamefont {Ma},
  \citenamefont {Giesler}, \citenamefont {Varma}, \citenamefont {Scheel},\ and\
  \citenamefont {Chen}}]{Ma_2021_SuperkickGWs}%
  \BibitemOpen
  \bibfield  {author} {\bibinfo {author} {\bibfnamefont {S.}~\bibnamefont
  {Ma}}, \bibinfo {author} {\bibfnamefont {M.}~\bibnamefont {Giesler}},
  \bibinfo {author} {\bibfnamefont {V.}~\bibnamefont {Varma}}, \bibinfo
  {author} {\bibfnamefont {M.~A.}\ \bibnamefont {Scheel}},\ and\ \bibinfo
  {author} {\bibfnamefont {Y.}~\bibnamefont {Chen}},\ }\bibfield  {title}
  {\bibinfo {title} {Universal features of gravitational waves emitted by
  superkick binary black hole systems},\ }\href
  {https://doi.org/10.1103/PhysRevD.104.084003} {\bibfield  {journal} {\bibinfo
   {journal} {Phys. Rev. D}\ }\textbf {\bibinfo {volume} {104}},\ \bibinfo
  {pages} {084003} (\bibinfo {year} {2021})},\ \Eprint
  {https://arxiv.org/abs/2107.04890} {arXiv:2107.04890 [gr-qc]} \BibitemShut
  {NoStop}%
\bibitem [{\citenamefont {Leong}\ \emph {et~al.}(2025)\citenamefont {Leong},
  \citenamefont {Tomé}, \citenamefont {Calderón~Bustillo}, \citenamefont {del
  Río},\ and\ \citenamefont {Sanchis-Gual}}]{Leong_2025_MirrorAsymmetry}%
  \BibitemOpen
  \bibfield  {author} {\bibinfo {author} {\bibfnamefont {S.~H.~W.}\
  \bibnamefont {Leong}}, \bibinfo {author} {\bibfnamefont {A.~F.}\ \bibnamefont
  {Tomé}}, \bibinfo {author} {\bibfnamefont {J.}~\bibnamefont
  {Calderón~Bustillo}}, \bibinfo {author} {\bibfnamefont {A.}~\bibnamefont
  {del Río}},\ and\ \bibinfo {author} {\bibfnamefont {N.}~\bibnamefont
  {Sanchis-Gual}},\ }\bibfield  {title} {\bibinfo {title} {Gravitational-wave
  signatures of the mirror asymmetry in binary black hole mergers:
  Measurability and correlation to gravitational-wave recoil},\ }\href
  {https://doi.org/10.1103/1nnp-w5w4} {\bibfield  {journal} {\bibinfo
  {journal} {Phys. Rev. D}\ }\textbf {\bibinfo {volume} {112}},\ \bibinfo
  {pages} {084078} (\bibinfo {year} {2025})},\ \Eprint
  {https://arxiv.org/abs/2501.11663} {arXiv:2501.11663 [gr-qc]} \BibitemShut
  {NoStop}%
\bibitem [{\citenamefont {Calderón~Bustillo}\ \emph
  {et~al.}(2025)\citenamefont {Calderón~Bustillo}, \citenamefont {del Rio},
  \citenamefont {Sanchis-Gual}, \citenamefont {Chandra},\ and\ \citenamefont
  {Leong}}]{Bustillo_2024_LVKMirrorAsymmetry}%
  \BibitemOpen
  \bibfield  {author} {\bibinfo {author} {\bibfnamefont {J.}~\bibnamefont
  {Calderón~Bustillo}}, \bibinfo {author} {\bibfnamefont {A.}~\bibnamefont
  {del Rio}}, \bibinfo {author} {\bibfnamefont {N.}~\bibnamefont
  {Sanchis-Gual}}, \bibinfo {author} {\bibfnamefont {K.}~\bibnamefont
  {Chandra}},\ and\ \bibinfo {author} {\bibfnamefont {S.~H.~W.}\ \bibnamefont
  {Leong}},\ }\bibfield  {title} {\bibinfo {title} {Testing mirror symmetry in
  the universe with ligo-virgo black-hole mergers},\ }\href
  {https://doi.org/10.1103/PhysRevLett.134.031402} {\bibfield  {journal}
  {\bibinfo  {journal} {Phys. Rev. Lett.}\ }\textbf {\bibinfo {volume} {134}},\
  \bibinfo {pages} {031402} (\bibinfo {year} {2025})},\ \Eprint
  {https://arxiv.org/abs/2402.09861} {arXiv:2402.09861 [gr-qc]} \BibitemShut
  {NoStop}%
\bibitem [{\citenamefont {{SXS
  Collaboration}}(2025{\natexlab{a}})}]{SXS_2025_SXSCatalogData300}%
  \BibitemOpen
  \bibfield  {author} {\bibinfo {author} {\bibnamefont {{SXS Collaboration}}},\
  }\href@noop {} {\bibinfo {title} {The sxs catalog of simulations v3.0.0}}
  (\bibinfo {year} {2025}{\natexlab{a}})\BibitemShut {NoStop}%
\bibitem [{\citenamefont {Boyle}\ \emph {et~al.}(2025)\citenamefont {Boyle},
  \citenamefont {Mitman}, \citenamefont {Scheel},\ and\ \citenamefont
  {Stein}}]{SXS_2025_SXSPackage}%
  \BibitemOpen
  \bibfield  {author} {\bibinfo {author} {\bibfnamefont {M.}~\bibnamefont
  {Boyle}}, \bibinfo {author} {\bibfnamefont {K.}~\bibnamefont {Mitman}},
  \bibinfo {author} {\bibfnamefont {M.}~\bibnamefont {Scheel}},\ and\ \bibinfo
  {author} {\bibfnamefont {L.}~\bibnamefont {Stein}},\ }\href
  {https://doi.org/10.5281/ZENODO.15547069} {\bibinfo {title} {The sxs
  package}} (\bibinfo {year} {2025})\BibitemShut {NoStop}%
\bibitem [{\citenamefont {{SXS
  Collaboration}}(2025{\natexlab{b}})}]{SXS_2025_SXSBBH1941}%
  \BibitemOpen
  \bibfield  {author} {\bibinfo {author} {\bibnamefont {{SXS Collaboration}}},\
  }\href {https://doi.org/10.26138/SXS:BBH:1941} {\bibinfo {title} {Binary
  black-hole simulation {SXS:BBH:1941}}} (\bibinfo {year}
  {2025}{\natexlab{b}})\BibitemShut {NoStop}%
\bibitem [{\citenamefont {{SXS
  Collaboration}}(2025{\natexlab{c}})}]{SXS_2025_SXSBBH3392}%
  \BibitemOpen
  \bibfield  {author} {\bibinfo {author} {\bibnamefont {{SXS Collaboration}}},\
  }\href {https://doi.org/10.26138/SXS:BBH:3392} {\bibinfo {title} {Binary
  black-hole simulation {SXS:BBH:3392}}} (\bibinfo {year}
  {2025}{\natexlab{c}})\BibitemShut {NoStop}%
\bibitem [{\citenamefont {Punturo}\ \emph {et~al.}(2010)\citenamefont {Punturo}
  \emph {et~al.}}]{Punturo_2010_ET}%
  \BibitemOpen
  \bibfield  {author} {\bibinfo {author} {\bibfnamefont {M.}~\bibnamefont
  {Punturo}} \emph {et~al.},\ }\bibfield  {title} {\bibinfo {title} {The
  einstein telescope: A third-generation gravitational wave observatory},\
  }\href {https://doi.org/10.1088/0264-9381/27/19/194002} {\bibfield  {journal}
  {\bibinfo  {journal} {Class. Quant. Grav.}\ }\textbf {\bibinfo {volume}
  {27}},\ \bibinfo {pages} {194002} (\bibinfo {year} {2010})}\BibitemShut
  {NoStop}%
\bibitem [{\citenamefont {Hild}\ \emph {et~al.}(2011)\citenamefont {Hild} \emph
  {et~al.}}]{Hild_2011_ETD}%
  \BibitemOpen
  \bibfield  {author} {\bibinfo {author} {\bibfnamefont {S.}~\bibnamefont
  {Hild}} \emph {et~al.},\ }\bibfield  {title} {\bibinfo {title} {Sensitivity
  studies for third-generation gravitational wave observatories},\ }\href
  {https://doi.org/10.1088/0264-9381/28/9/094013} {\bibfield  {journal}
  {\bibinfo  {journal} {Class. Quantum Gravity}\ }\textbf {\bibinfo {volume}
  {28}},\ \bibinfo {pages} {094013} (\bibinfo {year} {2011})},\ \Eprint
  {https://arxiv.org/abs/1012.0908} {arXiv:1012.0908 [gr-qc]} \BibitemShut
  {NoStop}%
\bibitem [{\citenamefont {Abac}\ \emph
  {et~al.}(2025{\natexlab{b}})\citenamefont {Abac} \emph
  {et~al.}}]{ET_2025_ETScience}%
  \BibitemOpen
  \bibfield  {author} {\bibinfo {author} {\bibfnamefont {A.}~\bibnamefont
  {Abac}} \emph {et~al.} (\bibinfo {collaboration} {ET}),\ }\bibfield  {title}
  {\bibinfo {title} {The science of the einstein telescope},\ }\href@noop {} {\
   (\bibinfo {year} {2025}{\natexlab{b}})},\ \Eprint
  {https://arxiv.org/abs/2503.12263} {arXiv:2503.12263 [gr-qc]} \BibitemShut
  {NoStop}%
\bibitem [{\citenamefont {Ashton}\ \emph {et~al.}(2019)\citenamefont {Ashton}
  \emph {et~al.}}]{Ashton_2019_Bilby}%
  \BibitemOpen
  \bibfield  {author} {\bibinfo {author} {\bibfnamefont {G.}~\bibnamefont
  {Ashton}} \emph {et~al.},\ }\bibfield  {title} {\bibinfo {title} {Bilby: A
  user-friendly bayesian inference library for gravitational-wave astronomy},\
  }\href {https://doi.org/10.3847/1538-4365/ab06fc} {\bibfield  {journal}
  {\bibinfo  {journal} {Astrophys. J.}\ }\textbf {\bibinfo {volume} {241}},\
  \bibinfo {pages} {27} (\bibinfo {year} {2019})},\ \Eprint
  {https://arxiv.org/abs/1811.02042} {arXiv:1811.02042 [astro-ph.IM]}
  \BibitemShut {NoStop}%
\bibitem [{\citenamefont {Speagle}(2020)}]{Speagle_2020_Dynesty}%
  \BibitemOpen
  \bibfield  {author} {\bibinfo {author} {\bibfnamefont {J.~S.}\ \bibnamefont
  {Speagle}},\ }\bibfield  {title} {\bibinfo {title} {Dynesty: a dynamic nested
  sampling package for estimating bayesian posteriors and evidences},\ }\href
  {https://doi.org/10.1093/mnras/staa278} {\bibfield  {journal} {\bibinfo
  {journal} {Mon. Not. R. Astron. Soc.}\ }\textbf {\bibinfo {volume} {493}},\
  \bibinfo {pages} {3132} (\bibinfo {year} {2020})},\ \Eprint
  {https://arxiv.org/abs/1904.02180} {arXiv:1904.02180 [astro-ph.IM]}
  \BibitemShut {NoStop}%
\bibitem [{\citenamefont {Hoy}\ and\ \citenamefont
  {Raymond}(2021)}]{Hoy_2021_PESummary}%
  \BibitemOpen
  \bibfield  {author} {\bibinfo {author} {\bibfnamefont {C.}~\bibnamefont
  {Hoy}}\ and\ \bibinfo {author} {\bibfnamefont {V.}~\bibnamefont {Raymond}},\
  }\bibfield  {title} {\bibinfo {title} {Pesummary: The code agnostic parameter
  estimation summary page builder},\ }\href
  {https://doi.org/https://doi.org/10.1016/j.softx.2021.100765} {\bibfield
  {journal} {\bibinfo  {journal} {SoftwareX}\ }\textbf {\bibinfo {volume}
  {15}},\ \bibinfo {pages} {100765} (\bibinfo {year} {2021})},\ \Eprint
  {https://arxiv.org/abs/2006.06639} {arXiv:2006.06639 [astro-ph.IM]}
  \BibitemShut {NoStop}%
\bibitem [{\citenamefont {Thompson}\ \emph {et~al.}(2025)\citenamefont
  {Thompson}, \citenamefont {Hoy}, \citenamefont {Fauchon-Jones},\ and\
  \citenamefont {Hannam}}]{Thompson_2025_indist}%
  \BibitemOpen
  \bibfield  {author} {\bibinfo {author} {\bibfnamefont {J.~E.}\ \bibnamefont
  {Thompson}}, \bibinfo {author} {\bibfnamefont {C.}~\bibnamefont {Hoy}},
  \bibinfo {author} {\bibfnamefont {E.}~\bibnamefont {Fauchon-Jones}},\ and\
  \bibinfo {author} {\bibfnamefont {M.}~\bibnamefont {Hannam}},\ }\bibfield
  {title} {\bibinfo {title} {On the use and interpretation of signal-model
  indistinguishability measures for gravitational-wave astronomy},\ }\href@noop
  {} {\bibfield  {journal} {\bibinfo  {journal} {Phys. Rev. D}\ }\textbf
  {\bibinfo {volume} {112}},\ \bibinfo {pages} {064011} (\bibinfo {year}
  {2025})},\ \Eprint {https://arxiv.org/abs/2506.10530} {arXiv:2506.10530
  [gr-qc]} \BibitemShut {NoStop}%
\bibitem [{\citenamefont {Scheel}\ \emph {et~al.}(2025)\citenamefont {Scheel}
  \emph {et~al.}}]{Scheel_2025_SXS3rdCatalog}%
  \BibitemOpen
  \bibfield  {author} {\bibinfo {author} {\bibfnamefont {M.~A.}\ \bibnamefont
  {Scheel}} \emph {et~al.},\ }\bibfield  {title} {\bibinfo {title} {The sxs
  collaboration's third catalog of binary black hole simulations},\ }\href
  {https://doi.org/10.1088/1361-6382/adfd34} {\bibfield  {journal} {\bibinfo
  {journal} {Class. Quantum Grav.}\ }\textbf {\bibinfo {volume} {42}},\
  \bibinfo {pages} {195017} (\bibinfo {year} {2025})},\ \Eprint
  {https://arxiv.org/abs/2505.13378} {arXiv:2505.13378 [gr-qc]} \BibitemShut
  {NoStop}%
\bibitem [{\citenamefont {{SXS
  Collaboration}}(2025{\natexlab{d}})}]{SXS_2025_SXSBBH1207}%
  \BibitemOpen
  \bibfield  {author} {\bibinfo {author} {\bibnamefont {{SXS Collaboration}}},\
  }\href {https://doi.org/10.26138/SXS:BBH:1207} {\bibinfo {title} {Binary
  black-hole simulation {SXS:BBH:1207}}} (\bibinfo {year}
  {2025}{\natexlab{d}})\BibitemShut {NoStop}%
\bibitem [{\citenamefont {Boyle}(2024)}]{Boyle_2024_PNjulia}%
  \BibitemOpen
  \bibfield  {author} {\bibinfo {author} {\bibfnamefont {M.}~\bibnamefont
  {Boyle}},\ }\href {https://github.com/moble/PostNewtonian.jl} {\bibinfo
  {title} {{PostNewtonian.jl}}} (\bibinfo {year} {2024})\BibitemShut {NoStop}%
\bibitem [{\citenamefont {Hamilton}\ \emph {et~al.}(2023)\citenamefont
  {Hamilton}, \citenamefont {London},\ and\ \citenamefont
  {Hannam}}]{Hamilton_2023_RDfreq}%
  \BibitemOpen
  \bibfield  {author} {\bibinfo {author} {\bibfnamefont {E.}~\bibnamefont
  {Hamilton}}, \bibinfo {author} {\bibfnamefont {L.}~\bibnamefont {London}},\
  and\ \bibinfo {author} {\bibfnamefont {M.}~\bibnamefont {Hannam}},\
  }\bibfield  {title} {\bibinfo {title} {Ringdown frequencies in black holes
  formed from precessing black-hole binaries},\ }\href
  {https://doi.org/10.1103/PhysRevD.107.104035} {\bibfield  {journal} {\bibinfo
   {journal} {Phys. Rev. D}\ }\textbf {\bibinfo {volume} {107}},\ \bibinfo
  {pages} {104035} (\bibinfo {year} {2023})},\ \Eprint
  {https://arxiv.org/abs/2301.06558} {arXiv:2301.06558 [gr-qc]} \BibitemShut
  {NoStop}%
\bibitem [{\citenamefont {Maga\~na Zertuche}\ and\ \citenamefont
  {Finch}(2025)}]{Zertuche_2025_qnmfits}%
  \BibitemOpen
  \bibfield  {author} {\bibinfo {author} {\bibfnamefont {L.}~\bibnamefont
  {Maga\~na Zertuche}}\ and\ \bibinfo {author} {\bibfnamefont {E.}~\bibnamefont
  {Finch}},\ }\href@noop {} {\bibinfo {title} {qnmfits}},\ \bibinfo
  {howpublished} {\url{https://github.com/sxs-collaboration/qnmfits}} (\bibinfo
  {year} {2025}),\ \bibinfo {note} {gitHub repository}\BibitemShut {NoStop}%
\bibitem [{\citenamefont {Stein}(2019)}]{Stein_2019_qnm}%
  \BibitemOpen
  \bibfield  {author} {\bibinfo {author} {\bibfnamefont {L.~C.}\ \bibnamefont
  {Stein}},\ }\bibfield  {title} {\bibinfo {title} {qnm: A python package for
  calculating kerr quasinormal modes, separation constants, and
  spherical-spheroidal mixing coefficients},\ }\href
  {https://doi.org/10.21105/joss.01683} {\bibfield  {journal} {\bibinfo
  {journal} {J. Open Source Softw.}\ }\textbf {\bibinfo {volume} {4}},\
  \bibinfo {pages} {1683} (\bibinfo {year} {2019})},\ \Eprint
  {https://arxiv.org/abs/1908.10377} {arXiv:1908.10377 [gr-qc]} \BibitemShut
  {NoStop}%
\bibitem [{\citenamefont {Cook}\ and\ \citenamefont
  {Zalutskiy}(2014)}]{Cook_2014_qnmfreqs}%
  \BibitemOpen
  \bibfield  {author} {\bibinfo {author} {\bibfnamefont {G.~B.}\ \bibnamefont
  {Cook}}\ and\ \bibinfo {author} {\bibfnamefont {M.}~\bibnamefont
  {Zalutskiy}},\ }\bibfield  {title} {\bibinfo {title} {Gravitational
  perturbations of the kerr geometry: High-accuracy study},\ }\href
  {https://doi.org/10.1103/physrevd.90.124021} {\bibfield  {journal} {\bibinfo
  {journal} {Phys. Rev. D}\ }\textbf {\bibinfo {volume} {90}},\ \bibinfo
  {pages} {124021} (\bibinfo {year} {2014})},\ \Eprint
  {https://arxiv.org/abs/1410.7698} {arXiv:1410.7698 [gr-qc]} \BibitemShut
  {NoStop}%
\bibitem [{\citenamefont {Cook}(2020)}]{Cook_2020_qnmmultimode}%
  \BibitemOpen
  \bibfield  {author} {\bibinfo {author} {\bibfnamefont {G.~B.}\ \bibnamefont
  {Cook}},\ }\bibfield  {title} {\bibinfo {title} {Aspects of multimode kerr
  ringdown fitting},\ }\href {https://doi.org/10.1103/PhysRevD.102.024027}
  {\bibfield  {journal} {\bibinfo  {journal} {Phys. Rev. D}\ }\textbf {\bibinfo
  {volume} {102}},\ \bibinfo {pages} {024027} (\bibinfo {year} {2020})},\
  \Eprint {https://arxiv.org/abs/2004.08347} {arXiv:2004.08347 [gr-qc]}
  \BibitemShut {NoStop}%
\bibitem [{\citenamefont {Mitman}\ \emph {et~al.}(2025)\citenamefont {Mitman},
  \citenamefont {Pretto}, \citenamefont {Siegel}, \citenamefont {Scheel},
  \citenamefont {Teukolsky}, \citenamefont {Boyle}, \citenamefont {Deppe},
  \citenamefont {Kidder}, \citenamefont {Moxon}, \citenamefont {Nelli},
  \citenamefont {Throwe},\ and\ \citenamefont {Vu}}]{Mitman_2025_VarPro}%
  \BibitemOpen
  \bibfield  {author} {\bibinfo {author} {\bibfnamefont {K.}~\bibnamefont
  {Mitman}}, \bibinfo {author} {\bibfnamefont {I.}~\bibnamefont {Pretto}},
  \bibinfo {author} {\bibfnamefont {H.}~\bibnamefont {Siegel}}, \bibinfo
  {author} {\bibfnamefont {M.~A.}\ \bibnamefont {Scheel}}, \bibinfo {author}
  {\bibfnamefont {S.~A.}\ \bibnamefont {Teukolsky}}, \bibinfo {author}
  {\bibfnamefont {M.}~\bibnamefont {Boyle}}, \bibinfo {author} {\bibfnamefont
  {N.}~\bibnamefont {Deppe}}, \bibinfo {author} {\bibfnamefont {L.~E.}\
  \bibnamefont {Kidder}}, \bibinfo {author} {\bibfnamefont {J.}~\bibnamefont
  {Moxon}}, \bibinfo {author} {\bibfnamefont {K.~C.}\ \bibnamefont {Nelli}},
  \bibinfo {author} {\bibfnamefont {W.}~\bibnamefont {Throwe}},\ and\ \bibinfo
  {author} {\bibfnamefont {N.~L.}\ \bibnamefont {Vu}},\ }\bibfield  {title}
  {\bibinfo {title} {Probing the ringdown perturbation in binary black hole
  coalescences with an improved quasi-normal mode extraction algorithm},\
  }\href@noop {} {\  (\bibinfo {year} {2025})},\ \Eprint
  {https://arxiv.org/abs/2503.09678} {arXiv:2503.09678 [gr-qc]} \BibitemShut
  {NoStop}%
\bibitem [{\citenamefont {Pacilio}\ \emph {et~al.}(2024)\citenamefont
  {Pacilio}, \citenamefont {Bhagwat}, \citenamefont {Nobili},\ and\
  \citenamefont {Gerosa}}]{Pacillio_2024_postmerger}%
  \BibitemOpen
  \bibfield  {author} {\bibinfo {author} {\bibfnamefont {C.}~\bibnamefont
  {Pacilio}}, \bibinfo {author} {\bibfnamefont {S.}~\bibnamefont {Bhagwat}},
  \bibinfo {author} {\bibfnamefont {F.}~\bibnamefont {Nobili}},\ and\ \bibinfo
  {author} {\bibfnamefont {D.}~\bibnamefont {Gerosa}},\ }\bibfield  {title}
  {\bibinfo {title} {Flexible mapping of ringdown amplitudes for nonprecessing
  binary black holes},\ }\href {https://doi.org/10.1103/PhysRevD.110.103037}
  {\bibfield  {journal} {\bibinfo  {journal} {Phys. Rev. D}\ }\textbf {\bibinfo
  {volume} {110}},\ \bibinfo {pages} {103037} (\bibinfo {year} {2024})},\
  \Eprint {https://arxiv.org/abs/2408.05276} {arXiv:2408.05276 [gr-qc]}
  \BibitemShut {NoStop}%
\bibitem [{\citenamefont {Maga\~na Zertuche}\ \emph {et~al.}(2025)\citenamefont
  {Maga\~na Zertuche} \emph {et~al.}}]{Zertuche_2025_NRSur2dq8RD}%
  \BibitemOpen
  \bibfield  {author} {\bibinfo {author} {\bibfnamefont {L.}~\bibnamefont
  {Maga\~na Zertuche}} \emph {et~al.},\ }\bibfield  {title} {\bibinfo {title}
  {High-precision ringdown surrogate model for nonprecessing binary black
  holes},\ }\href {https://doi.org/10.1103/q7sy-g3kl} {\bibfield  {journal}
  {\bibinfo  {journal} {Phys. Rev. D}\ }\textbf {\bibinfo {volume} {112}},\
  \bibinfo {pages} {024077} (\bibinfo {year} {2025})},\ \Eprint
  {https://arxiv.org/abs/2408.05300} {arXiv:2408.05300 [gr-qc]} \BibitemShut
  {NoStop}%
\bibitem [{\citenamefont {Zhu}\ \emph {et~al.}(2025{\natexlab{a}})\citenamefont
  {Zhu} \emph {et~al.}}]{Zhu_2025_BHSpectroscopy}%
  \BibitemOpen
  \bibfield  {author} {\bibinfo {author} {\bibfnamefont {H.}~\bibnamefont
  {Zhu}} \emph {et~al.},\ }\bibfield  {title} {\bibinfo {title} {Black hole
  spectroscopy for precessing binary black hole coalescences},\ }\href
  {https://doi.org/10.1103/PhysRevD.111.064052} {\bibfield  {journal} {\bibinfo
   {journal} {Phys. Rev. D}\ }\textbf {\bibinfo {volume} {111}},\ \bibinfo
  {pages} {064052} (\bibinfo {year} {2025}{\natexlab{a}})},\ \Eprint
  {https://arxiv.org/abs/2312.08588} {arXiv:2312.08588 [gr-qc]} \BibitemShut
  {NoStop}%
\bibitem [{\citenamefont {Zhu}\ \emph {et~al.}(2025{\natexlab{b}})\citenamefont
  {Zhu}, \citenamefont {Siegel},\ and\ \citenamefont {Mitman}}]{Zhu_2025_git}%
  \BibitemOpen
  \bibfield  {author} {\bibinfo {author} {\bibfnamefont {H.}~\bibnamefont
  {Zhu}}, \bibinfo {author} {\bibfnamefont {H.}~\bibnamefont {Siegel}},\ and\
  \bibinfo {author} {\bibfnamefont {K.}~\bibnamefont {Mitman}},\ }\href@noop {}
  {\bibinfo {title} {precession-ringdown}},\ \bibinfo {howpublished}
  {\url{https://github.com/HengruiZhu99/precession_ringdown}} (\bibinfo {year}
  {2025}{\natexlab{b}}),\ \bibinfo {note} {gitHub repository}\BibitemShut
  {NoStop}%
\bibitem [{\citenamefont {Nobili}\ \emph {et~al.}(2025)\citenamefont {Nobili},
  \citenamefont {Bhagwat}, \citenamefont {Pacilio},\ and\ \citenamefont
  {Gerosa}}]{Nobili_2025_RDAmpl}%
  \BibitemOpen
  \bibfield  {author} {\bibinfo {author} {\bibfnamefont {F.}~\bibnamefont
  {Nobili}}, \bibinfo {author} {\bibfnamefont {S.}~\bibnamefont {Bhagwat}},
  \bibinfo {author} {\bibfnamefont {C.}~\bibnamefont {Pacilio}},\ and\ \bibinfo
  {author} {\bibfnamefont {D.}~\bibnamefont {Gerosa}},\ }\bibfield  {title}
  {\bibinfo {title} {Ringdown mode amplitudes of precessing binary black
  holes},\ }\href@noop {} {\  (\bibinfo {year} {2025})},\ \Eprint
  {https://arxiv.org/abs/2504.17021} {arXiv:2504.17021 [gr-qc]} \BibitemShut
  {NoStop}%
\bibitem [{\citenamefont {Harris}\ \emph {et~al.}(2020)\citenamefont {Harris}
  \emph {et~al.}}]{Harris_2020_NumPy}%
  \BibitemOpen
  \bibfield  {author} {\bibinfo {author} {\bibfnamefont {C.~R.}\ \bibnamefont
  {Harris}} \emph {et~al.},\ }\bibfield  {title} {\bibinfo {title} {Array
  programming with numpy},\ }\href {https://doi.org/10.1038/s41586-020-2649-2}
  {\bibfield  {journal} {\bibinfo  {journal} {Nature}\ }\textbf {\bibinfo
  {volume} {585}},\ \bibinfo {pages} {357} (\bibinfo {year} {2020})},\ \Eprint
  {https://arxiv.org/abs/2006.10256} {arXiv:2006.10256 [cs.MS]} \BibitemShut
  {NoStop}%
\bibitem [{\citenamefont {Hunter}(2007)}]{Hunter_2007_Matplotlib}%
  \BibitemOpen
  \bibfield  {author} {\bibinfo {author} {\bibfnamefont {J.~D.}\ \bibnamefont
  {Hunter}},\ }\bibfield  {title} {\bibinfo {title} {Matplotlib: A 2d graphics
  environment},\ }\href {https://doi.org/10.1109/MCSE.2007.55} {\bibfield
  {journal} {\bibinfo  {journal} {Comput. Sci. Eng.}\ }\textbf {\bibinfo
  {volume} {9}},\ \bibinfo {pages} {90} (\bibinfo {year} {2007})}\BibitemShut
  {NoStop}%
\bibitem [{\citenamefont {Virtanen}\ \emph {et~al.}(2020)\citenamefont
  {Virtanen} \emph {et~al.}}]{Virtanen_2020_SciPy}%
  \BibitemOpen
  \bibfield  {author} {\bibinfo {author} {\bibfnamefont {P.}~\bibnamefont
  {Virtanen}} \emph {et~al.},\ }\bibfield  {title} {\bibinfo {title} {Scipy
  1.0: fundamental algorithms for scientific computing in python},\ }\href
  {https://doi.org/10.1038/s41592-019-0686-2} {\bibfield  {journal} {\bibinfo
  {journal} {Nat. Methods}\ }\textbf {\bibinfo {volume} {17}},\ \bibinfo
  {pages} {261} (\bibinfo {year} {2020})},\ \Eprint
  {https://arxiv.org/abs/1907.10121} {arXiv:1907.10121 [cs.MS]} \BibitemShut
  {NoStop}%
\bibitem [{\citenamefont {{LIGO Scientific Collaboration}}\ \emph
  {et~al.}(2018)\citenamefont {{LIGO Scientific Collaboration}}, \citenamefont
  {{Virgo Collaboration}},\ and\ \citenamefont {{KAGRA
  Collaboration}}}]{LVK_2018_LALSuite}%
  \BibitemOpen
  \bibfield  {author} {\bibinfo {author} {\bibnamefont {{LIGO Scientific
  Collaboration}}}, \bibinfo {author} {\bibnamefont {{Virgo Collaboration}}},\
  and\ \bibinfo {author} {\bibnamefont {{KAGRA Collaboration}}},\ }\href
  {https://doi.org/10.7935/GT1W-FZ16} {\bibinfo {title} {{LVK} {A}lgorithm
  {L}ibrary - {LALS}uite}},\ \bibinfo {howpublished} {Free software (GPL)}
  (\bibinfo {year} {2018})\BibitemShut {NoStop}%
\bibitem [{\citenamefont {Field}\ \emph {et~al.}(2014)\citenamefont {Field},
  \citenamefont {Galley}, \citenamefont {Hesthaven}, \citenamefont {Kaye},\
  and\ \citenamefont {Tiglio}}]{Field_2013_GWSurrogate}%
  \BibitemOpen
  \bibfield  {author} {\bibinfo {author} {\bibfnamefont {S.~E.}\ \bibnamefont
  {Field}}, \bibinfo {author} {\bibfnamefont {C.~R.}\ \bibnamefont {Galley}},
  \bibinfo {author} {\bibfnamefont {J.~S.}\ \bibnamefont {Hesthaven}}, \bibinfo
  {author} {\bibfnamefont {J.}~\bibnamefont {Kaye}},\ and\ \bibinfo {author}
  {\bibfnamefont {M.}~\bibnamefont {Tiglio}},\ }\bibfield  {title} {\bibinfo
  {title} {Fast prediction and evaluation of gravitational waveforms using
  surrogate models},\ }\href {https://doi.org/10.1103/PhysRevX.4.031006}
  {\bibfield  {journal} {\bibinfo  {journal} {Phys. Rev. X}\ }\textbf {\bibinfo
  {volume} {4}},\ \bibinfo {pages} {031006} (\bibinfo {year} {2014})},\ \Eprint
  {https://arxiv.org/abs/1308.3565} {arXiv:1308.3565 [gr-qc]} \BibitemShut
  {NoStop}%
\end{thebibliography}%
\end{document}